\documentclass[a4paper,fleqn,usenatbib]{mnras}

\usepackage[T1]{fontenc}
\usepackage{ae,aecompl}

\usepackage{graphicx}	
\usepackage{amsmath}	
\usepackage{amssymb}	

\newcommand{\Msun}{$\mathrm{M}\odot$}
\newcommand{\Rsun}{$\mathrm{R}\odot$}

\title[Fast PISNe]
{Fast evolving pair-instability supernova models: \\ 
Evolution, explosion, light curves}

\author[A. Kozyreva et al.]
{Alexandra~Kozyreva$^{1,2}$\thanks{E-mail: a.kozyreva@keele.ac.uk},
Matthew~Gilmer$^{3}$, Raphael~Hirschi$^{2,4}$, 
\newauthor
Carla~Fr\"ohlich$^{3}$, Sergey~Blinnikov$^{4,5,6}$, Ryan~T.~Wollaeger$^{7}$, 
\newauthor
Ulrich~M.~Noebauer$^8$, Daniel R. van Rossum$^{9}$, Alexander~Heger$^{10,11,12}$, 
\newauthor
Wesley~P.~Even$^{7}$, Roni~Waldman$^{13}$, Alexey~Tolstov$^{4}$, 
\newauthor
Emmanouil~Chatzopoulos$^{14}$, Elena~Sorokina$^{4,5,15}$
\\
$^{1}$The Raymond and Beverly Sackler School of Physics and Astronomy, Tel Aviv University, Tel Aviv, 69978, Israel \\
$^{2}$Astrophysics group, Keele University, Keele, Staffordshire, ST5 5BG, UK \\
$^{3}$Department of Physics, North Carolina State University, Raleigh, NC 27695-8202, USA\\
$^{4}$Kavli Institute for the Physics and Mathematics of the Universe (WPI),
The University of Tokyo Institutes for Advanced Study, \\
The University of Tokyo, 5-1-5 Kashiwanoha, Kashiwa, Chiba 277-8583, Japan\\
$^{5}$ITEP (Kurchatov Institute), Moscow, 117218, Russia \\
$^{6}$VNIIA, Moscow, 127055, Russia\\  
$^{7}$Center for Theoretical Astrophysics/CCS-2, Los Alamos National Laboratory, Los Alamos, NM 87544, USA\\
$^{8}$Max-Planck-Institut f\"ur Astrophysik, Karl-Schwarzschild-Stra$\beta$e 1, D-85748 Garching, Germany\\
$^{9}$Flash Center for Computational Science, University of Chicago, Chicago, IL 60637, USA\\
$^{10}$Monash Centre for Astrophysics, School of Physics and Astronomy, Monash University, Victoria, 3800, Australia\\
$^{11}$Center for Nuclear Astrophysics, Department of Physics and Astronomy, Shanghai Jiao-Tong University, Shanghai 200240, P. R. China\\
$^{12}$School of Physics and Astronomy, University of Minnesota, Minneapolis, MN 55455, USA\\
$^{13}$Racah Institute of Physics, The Hebrew University, Jerusalem 91904, Israel\\
$^{14}$Department of Physics and Astronomy, Louisiana State University, Baton Rouge, LA 70803-4001, USA\\
$^{15}$Sternberg Astronomical Institute, M.V.~Lomonosov Moscow State
University, Moscow, 119991, Russia 
}

\date{Accepted XXX. Received YYY; in original form ZZZ}

\pubyear{2016}

\begin{document}
\label{firstpage}
\pagerange{\pageref{firstpage}--\pageref{lastpage}}
\maketitle

\begin{abstract}
With an increasing number of superluminous supernovae (SLSNe) discovered the question of
their origin remains open and causes heated debates in the
supernova community.  Currently, there are three proposed mechanisms for SLSNe: (1)
pair-instability supernovae (PISN), (2) magnetar-driven
supernovae, and (3) models in which the supernova ejecta interacts with a
circumstellar material ejected before the explosion.  
Based on current observations of SLSNe,
the PISN origin has been disfavoured
for a number of reasons. Many PISN models provide overly broad light curves and
too reddened spectra, because of massive ejecta and a high amount of nickel.  
In the current study we re-examine PISN properties using
progenitor models computed with the \texttt{GENEC} code.  
We calculate supernova explosions with \texttt{FLASH} and light
curve evolution with the radiation hydrodynamics code \texttt{STELLA}.  
We find that high-mass models (200~\Msun{} and 250~\Msun{}) at relatively high 
metallicity ({\it{Z}}\,=\,0.001) do not retain hydrogen in the outer layers
and produce relatively fast evolving PISNe Type~I and
might be suitable to explain some SLSNe.  We also investigate uncertainties in
light curve modelling due to codes, opacities, the nickel-bubble effect and 
progenitor structure and composition.



\end{abstract}




\section[Introduction]{Introduction}
\label{sect:intro}

The evolution of very massive stars\footnote{According to the
analysis and definition by
\citet{1982sscr.conf..303B} and \citet{1984ApJ...277..445C}, a very massive star is
a star with initial mass above approximately 100~\Msun{} and below
several $10^{\,4}$~\Msun{}.} at zero metallicity, i.e. without mass loss, with initial mass between
approximately 140~\Msun{} and 260~\Msun{}, is more or less clear.  Following the
sequence of hydrostatic hydrogen, helium, carbon and neon burning, the
hydrodynamical instability develops due to electron--positron pair
creation caused by dominating radiation pressure
\citep[][and others]{1967SvA....10..604B,1967PhRvL..18..379B,1967ApJ...148..803R,1968Ap&SS...2...96F}. 
Subsequently, oxygen and
silicon burn explosively.  If the nuclear burning energy released exceeds
the binding energy of the star, the star blows up in an explosion --
a pair-instability supernova (hereafter PISN).   The amount of radioactive
nickel generated during the explosion phase may be as high as 55~\Msun{}
\citep{2002ApJ...567..532H} resulting in a very bright supernova event.  
Nevertheless, the major uncertainties in the evolution of very massive
stars are the mass-loss prescriptions and the treatment of convection 
\citep{2015ASSL..412...77V,2015ASSL..412..199W}.

Observationally, astrophysicists have clear confirmation of the
existence of very massive stars 
in nearby galaxies \citep[see e.g.][]{2010MNRAS.408..731C,2014ApJ...780..117S}.  
In fact, it is hard to measure the mass of an individual star, as many
massive stars are born in tight clusters
(\citealt{2003ARA&A..41...57L,2007ARA&A..45..481Z}, see also discussion in
\citealt{2014A&A...566A...6H}).  With the approaching launch of the James Webb
Space Telescope, it will
still be difficult to resolve individual stars \citep{2013MNRAS.429.3658R}.  
Even if it is difficult to catch glimpse of
these rare very massive stars, their powerful explosions, i.e. supernovae,
may be detectable up to very high redshifts \citep{2013ApJ...762L...6W}.  
Fitting the supernova observations with
the theoretical simulations primarily helps in understanding the 
evolution and explosion of these very massive star populations.

Modern large survey telescopes lead to the detection of hundreds of
supernovae each year \citep{2013PASP..125..749G}.  A small fraction of these supernovae reach a
significantly higher peak luminosity than an average supernova
\citep{2011MNRAS.412.1419L,2011MNRAS.412.1473L,2011Natur.474..487Q,2012Sci...337..927G,2014MNRAS.444.2096N,2014AJ....147..118R}.  
One of the possible mechanisms for these superluminous explosions is PISNe powered by
radioactive nickel decay.


Recent studies clearly show that metal--free ($Z=0$) or almost metal-free
($Z=10^{\,-4} Z_\odot$) PISN models retain a very massive hydrogen-rich
envelope because of an absence of mass-loss or a very low mass-loss rate
\citep{2011ApJ...734..102K,2013MNRAS.428.3227D}.  
There is a large uncertainty because mass-loss rates at low
metallicity are extrapolated from rates derived for considerably 
higher metallicity \citep{2007A&A...461..571H,2015ASSL..412...77V}.  
These metal-free progenitors originate in 
a low-metallicity environment, i.e. in the early Universe
\citep{2005ApJ...633.1031S,2012MNRAS.422.2701P,2013ApJ...777..110W}.
As a consequence, these massive PISNe display very broad light curves
\citep{2011ApJ...734..102K,2013MNRAS.428.3227D}.  The large
amount of radioactive nickel powers the light curve at maximum and makes it
brighter and broader for higher nickel mass \citep{2016MNRAS.tmpL..20K}.  
Even assuming a hydrogen--free ejecta, the resulting light curves are still too broad to be
considered viable candidates for most SLSNe
\citep{2011ApJ...734..102K,2015ApJ...799...18C}.  
A hydrogen-free PISN progenitor may originate from the evolution
of a hydrogen-rich star which lost its hydrogen during hydrogen and helium
core burning via stellar winds, pulsations or binary interactions
\citep{2001ApJ...550..890B}.  
In addition, a number of studies show that their spectra are too red, both at earlier
times \citep{2012MNRAS.426L..76D,2013MNRAS.428.3227D,2015ApJ...799...18C} and 
during the nebular phase \citep{2016MNRAS.455.3207J}, this makes PISNe
with or without hydrogen inadequate for explaining blue SLSN spectra
\citep{2014MNRAS.444.2096N,2015MNRAS.452.3869N}.

The situation is different for PISN progenitors at higher metallicity, $Z\sim\,0.001$.  
If stars retain hydrogen in their atmosphere, the light curves are still
broad, but not as broad as their metal--free siblings \citep{2014A&A...565A..70K}.  
Therefore, they are good candidates for
explaining of at least some slowly evolving SLSNe like SN\,2007bi.  
Although, the colour temperatures for these PISNe hardly matches
the majority of SLSNe, they are reasonably close to the colour temperatures for
slowly evolving SLSNe.  
It may happen that very massive stars at non-zero metallicity 
($Z=10^{\,-3}$ to $Z=2 \times 10^{\,-3}$, and higher, up to the PISN metallicity threshold
$Z=6 \times 10^{\,-3}$) never retain hydrogen \citep{2013MNRAS.433.1114Y,2015ASSL..412..157H}.  
The stellar evolution simulations show that stars quickly lose their hydrogen atmosphere and 
in the most extreme cases
also lose most of their helium layer, leaving a 2--3~\Msun{} shallow helium envelope.  
This is mainly caused by the mass-loss rate which is higher for
higher metallicity \citep[see][for more details]{2013MNRAS.433.1114Y}.  
Because of the low helium abundance and absence of hydrogen, 
nothing prevents the recombination front from rapidly moving through the outer layers
and reach the cloud of diffusing photons produced by the decay of the nickel and cobalt.  
These very massive stars might result in a faster evolving PISNe.

In this study we consider calculations for 
PISN progenitors which lost all hydrogen and a large fraction of helium.  
We analyse our numerical results in the context of the SLSN PTF12dam.  

We describe our models in Section~\ref{sect:method}, present 
the resulting light curves and photospheric evolution in
Section~\ref{sect:results}.  In Section~\ref{sect:compare}, 
we discuss the results in the context of SLSNe.  Comparative analysis is
done in Section~\ref{sect:code2code}.  
We conclude our study in Section~\ref{sect:conclusions}.

\section[Input models and light curves modelling]{Input models and light curve modelling}
\label{sect:method}

\subsection[Stellar evolution models]{Stellar evolution models}
\label{subsect:evol}

\begin{table*}
\caption[PISN models]
{Characteristics of the PISN models.  
All masses are in solar masses (H for hydrogen, He for
helium, CO-core -- for the carbon-oxygen core, defined as the mass
coordinate where $\mathrm{X}_\mathrm{C}+\mathrm{X}_\mathrm{O}=0.5$,
$^{56}$Ni for radioactive nickel).  For helium, the numbers in
parenthesis stand for helium mass only in the outer layers.  
E$_\mathrm{kin}$ is the kinetic
energy at infinity in Bethe (B), i.e. in $10^{\,51}$~erg.  M$_\mathrm{above\,Ni}$
stands for mass above the region containing $^{56}$Ni, i.e. the shell where $^{56}$Ni mass
fraction turns below $10^{\,-4}$. v$_\mathrm{Ni}$ stands for the velocity of
this shell.  We include the helium PISN model He130
\citep{2011ApJ...734..102K} and the hydrogen-rich PISN model 250M \citep{2014A&A...565A..70K} for comparison.
}
\label{table:models}
\begin{center}
\begin{tabular}{r|c|c|c|c|c|c|c|c|c|c|c|c}
model  & M$_\mathrm{fin}$ & Z & R & H  & H$_\mathrm{\,surf}$ & He & He$_\mathrm{\,surf}$ & CO-core   & $^{56}$Ni &
M$_\mathrm{above\,Ni}$ & 
v$_\mathrm{Ni}$ & E$_\mathrm{kin}$  \\
name   & [\Msun{}] &  & [\Rsun{}]& [\Msun{}]&& [\Msun{}]&& [\Msun{}] & [\Msun{}] & [\Msun{}] &
[1000\,km\,s$^{\,-1}$] & [B] \\
\hline
P200   & 110 & 0.001 & 81  &0.01& 0.05 & 9     & 0.94 & 100 & 12 & 95 &3.2 &53 \\
P250   & 127 & 0.001 & 2   &0   & 0    & 2.6(2)& 0.34 & 116 & 34 & 49 &7.5 &86 \\
He130  & 130 & 0     & 7   &0   & 0    & 2.8(1.65)& 1 & 121 & 40 & 64 &8   &90 \\
250M   & 169 & 0.001 & 745 &10  & 0.27 & 48    & 0.72 & 110 & 19 & 115&5.3 &48 \\
\end{tabular}
\end{center}
\end{table*}

Our main input models are the following: non--rotating 200~\Msun{} and
250~\Msun{} stars at metallicity $Z=10^{\,-3}$ (hereafter, P200 and P250,
see Table~\ref{table:models}).  The evolution during
hydrogen, helium, carbon core burning is
computed with the stellar evolution code \texttt{GENEC}
\citep{2012A&A...537A.146E,2013MNRAS.433.1114Y}.  The details of the physical ingredients of the
models are as described by \citet{2012A&A...537A.146E}.  We list the main features here:

\begin{itemize}
\item The initial abundances for these models are adapted from \citet{APS05} except for the neon abundances adopted from 
\citet{Cunha06}, considering enhanced $\alpha$--element abundances and a total metallicity, $Z=10^{\,-3}$. 
\item Nuclear reaction rates are generated by NetGen tools where they take most of the data from NACRE
\citep{1999NuPhA.656....3A}.  The current NACRE data has been redetermined and updated and 
some of the comparison to NACRE values and a short description of the 
effects on stellar evolution has been described in \citet{2012A&A...537A.146E}.
\item Neutrino energy loss in plasma, including pair and photo-neutrino
processes is taken from \citet{Itoh89} and \citet{Itoh96}.
\item Opacity is taken from OPAL \citep{1996ApJ...464..943I} and complemented with low 
temperature opacities from \citep{AF05} adapted for the high neon abundance.
\item The convective core is extended with an overshoot parameter
$\mathrm{d_{over}}/H_P$ = 0.10 starting from the Schwarzschild limit.
\item Since models calculated are $>100 M_\odot$, the outer convective zone is treated according to mixing
length theory, using $\alpha_\mathrm{MLT}$ = 1.0.  This is because, for the most luminous models, the
turbulent pressure and
acoustic flux need to be included in the treatment of the envelope.  The choice of outer convective zone 
for different initial mass has been described in detail in \citet{2012A&A...537A.146E}.
\item We adopted mass loss for hot O~stars from
\citet{2001A&A...369..574V}.  When the models reach the Wolf-Rayet (WR) transition,
i.e. hydrogen surface abundance drops below 0.3
we adopted the mass-loss rate of WR from \citet{2000A&A...360..227N} or \citet{GH08} depending 
on which effective temperature is reached by
the models.  For the temperature domains not covered by \citet{2001A&A...369..574V} and
\citet{2000A&A...360..227N} or \citet{GH08}, the mass loss prescription from \citet{deJager88} is used.
\end{itemize}

\begin{figure*}
\centering
\begin{tabular}{cc}
\includegraphics[width=0.5\textwidth,clip=]{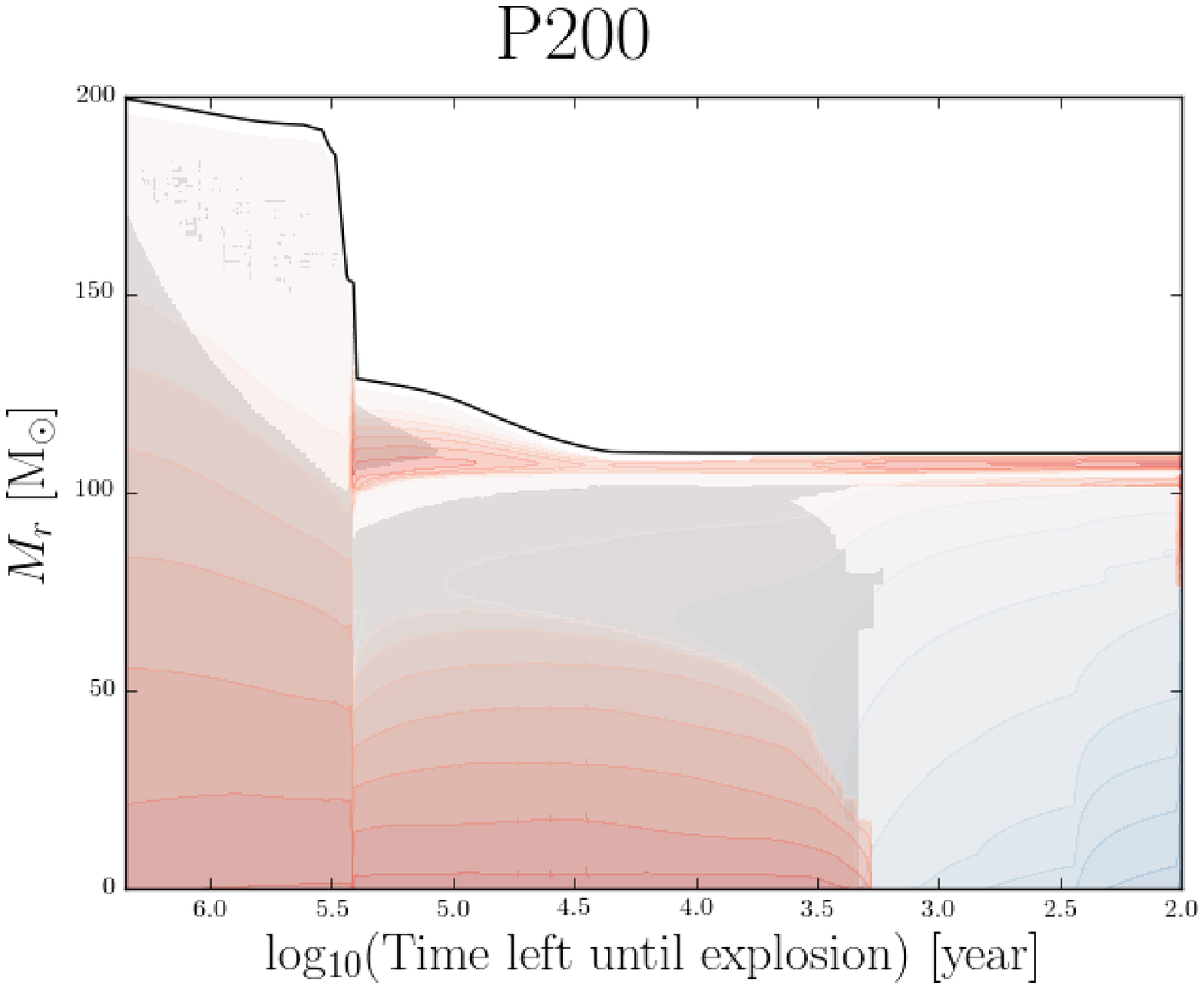} &
\includegraphics[width=0.5\textwidth,clip=]{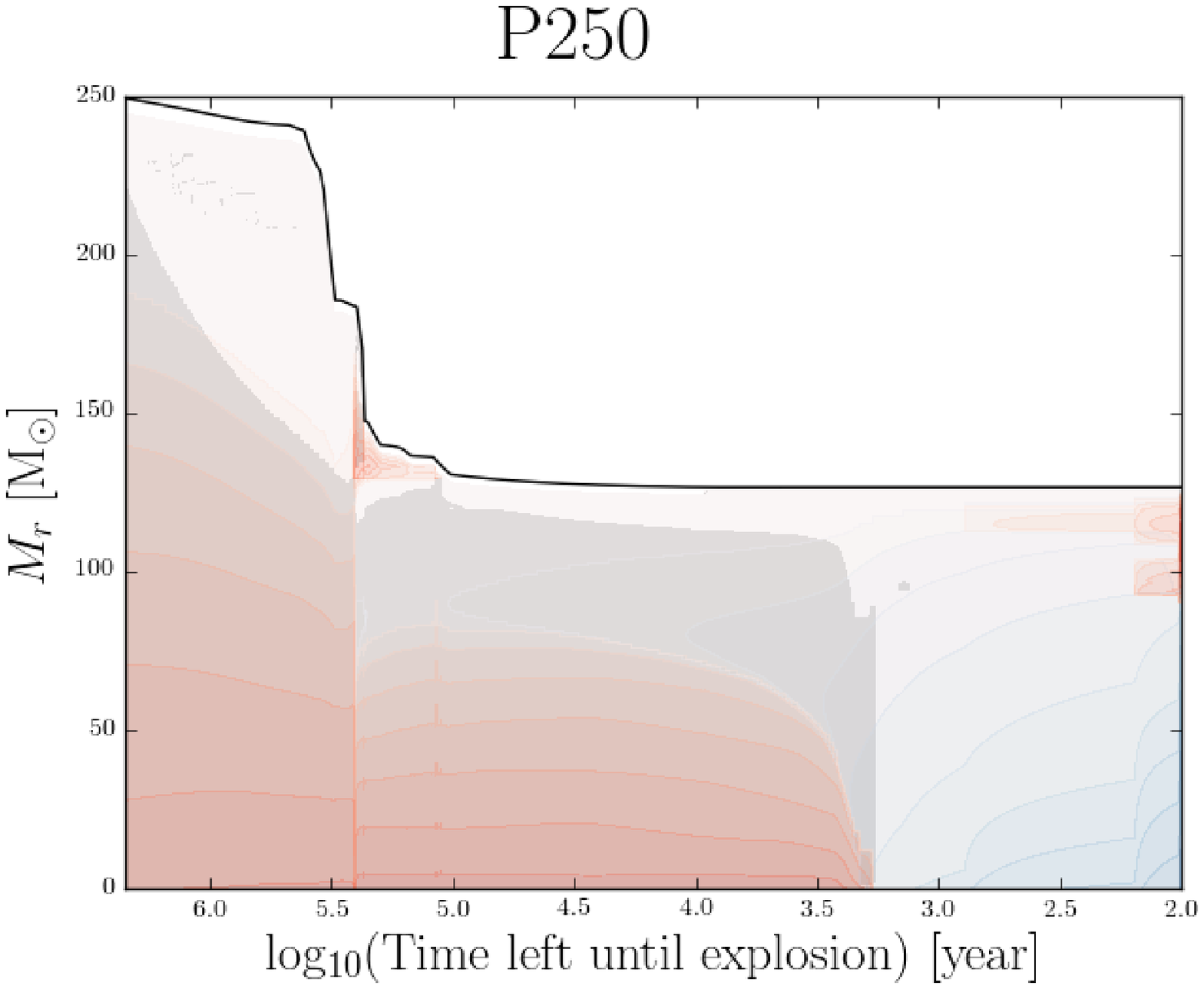} 
\end{tabular}
\caption{Structure evolution (aka Kippenhahn) diagrams for the 200~\Msun{}
(P200, {\it left}) and 250~\Msun{} (P250, {\it right}) as a function of age.  
The grey zones represent the convective regions.   
The top solid line corresponds to the total mass.  
Reddish area indicates the regions where energy is released
via nuclear burning, and bluish area indicates cooling via neutrino
losses.}
\label{figure:Kipp}
\end{figure*}

The evolution of the two models is shown in Fig.~\ref{figure:Kipp}.  Both models experience strong mass loss both just 
before and just after the main sequence.  This is due to the models reaching low enough temperatures to first 
reach the bi-stability limit \citep{2001A&A...369..574V} and then the limit of the domain of validity of the
\citet{2001A&A...369..574V} prescriptions.  
At this point, the code switches to the \citet{deJager88} mass loss, which is an uncertain empirical prescription 
including strong mass-loss linked to the luminous variable phase
\citep{1993MNRAS.263..375G,1994PASP..106.1025H}.  
After most of the hydrogen-rich envelope is lost, the surface layers contract and the models enter the Wolf-Rayet 
phase, during which mass-loss rates become relatively modest at $Z=10^{\,-3}$ (around $10^{-4}$ solar masses per year 
as opposed to up to $10^{-2.5}$ solar masses per year during the LBV phase).  While model P200 retains a small amount of 
hydrogen near its surface (0.05~\Msun{}), model P250 loses all of its hydrogen and most of its helium.

\subsection[PI explosion]{Pair-instability explosion}
\label{subsect:expl}

Near the end of carbon burning a fraction of the oxygen core undergoes
a dynamical instability as dominating radiation pressure allows the production of
electron--positron pairs.  
Even though the equation of state of both radiation-dominated plasma and a
mixture of plasma+radiation+pairs have an adiabatic index close to 4/3 ($P\sim\rho^{\,4/3}$),
there is an offset between them.  The {\emph{phase transition}} between
``radiation'' and ``radiation+pairs'' thus causes the effective adiabatic index
to drop below its equilibrium value of 4/3 
\citep[see Figure~32 on page~237 in][]{1971reas.book.....Z,1981ppse.book.....Z,1996ApJS..106..171B}.  However,
\texttt{GENEC} is not capable of following this instability, because the
equation of state implemented into \texttt{GENEC} does not include
electron--positron pair pressure.  Therefore, 
models were mapped into the hydrodynamical code
\texttt{FLASH}\footnote{\url{http://flash.uchicago.edu/site/flashcode/}} at this evolutionary stage 
\citep[version~4.3,][]{2000ApJS..131..273F,2009arXiv0903.4875D,2013ApJ...776..129C,2015ApJ...799...18C}.  
For the \texttt{FLASH} simulations, we used the Helmholtz equation-of-state \citep{2000ApJS..126..501T}
which includes pressure contributions from electron-positron pairs.  The
nuclear burning is calculated according to the 19-isotope reaction network
Aprox19\footnote{\url{http://cococubed.asu.edu/code_pages/net_torch.shtml}}, which includes $\alpha$--chain and
heavy-ion reactions as well as photo-disintegration and nucleon captures
between the isotopes, $^{52}$Fe, $^{54}$Fe, and $^{56}$Ni.  
The energy generation rates are calculated from the derivatives of abundances.  
This allows us to calculate the explosive nuclear burning coupled with hydrodynamics.  
The nuclear timescale becomes comparable to the dynamical
timescale at the end of neon core burning and throughout the explosive phase.

All of our \texttt{FLASH} simulations were carried out in
spherical symmetry with the new directionally-unsplit hydrodynamics solver
\citep{2009ASPC..406..243L} using the third-order piece-wise parabolic method 
\citep[PPM,][]{1984JCoPh..54..115W,1984JCoPh..54..174C,1989nuas.conf..100F}.  For
both our models, the core (within $5\times10^{\,10}$~cm for P250 or within
$4.167\times10^{\,10}$~cm for P200) was mapped first into \texttt{FLASH} and evolved through collapse
to the onset of explosion, and until all nuclear burning was completed.  The
initial envelope was then appended onto the exploding core and mapped back
into \texttt{FLASH} to follow shock burning up until the moment before shock
break-out.  To achieve convergence in the explosion properties with
resolution, we performed a series of simulations varying the maximum
refinement level as well as the refinement criteria, while the minimum
resolution remained constant at $4.4\times10^{\,8}$~cm.  The maximum resolution ranged
from $1.1\times10^{\,8}$~cm to $6.9\times10^{\,6}$~cm and the refinement criteria were modified
in order to allow the various maximum refinement levels to be reached in the
central regions during the explosive burning phase.  Variations in the total
nickel-56 yield for the above range of maximum resolutions were at the 17\%
level.  The simulations presented here used a maximum refinement of
$6.5\times10^{\,7}$~cm and produced 12~\Msun{} and 34~\Msun{} of
nickel-56 for models P200 and P250, respectively.  We will present details
of the \texttt{FLASH} simulations in the forthcoming paper (Gilmer~et~al. in preparation).  
The collapse phase and explosion
phase are computed without any special non-physical assumptions.  The collapse is
caused naturally by a hydrodynamical instability, since \texttt{FLASH}
properly treats the inclusion of pairs in the equation of state.   The explosion
is driven by the energy deposition from oxygen and silicon nuclear burning
followed by the \texttt{FLASH} nuclear network.  

\begin{figure*}
\centering
\includegraphics[width=0.5\textwidth]{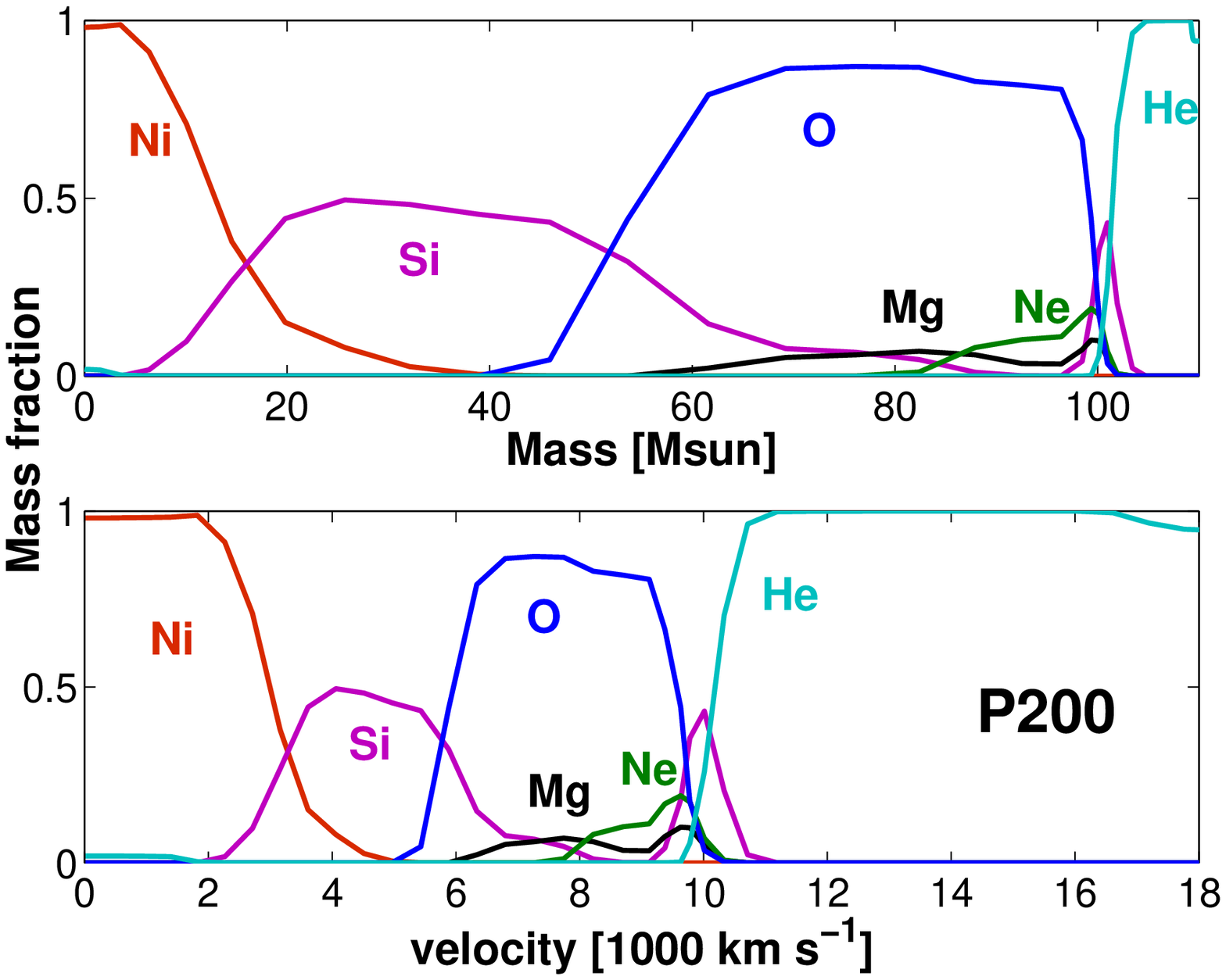}~~
\includegraphics[width=0.5\textwidth]{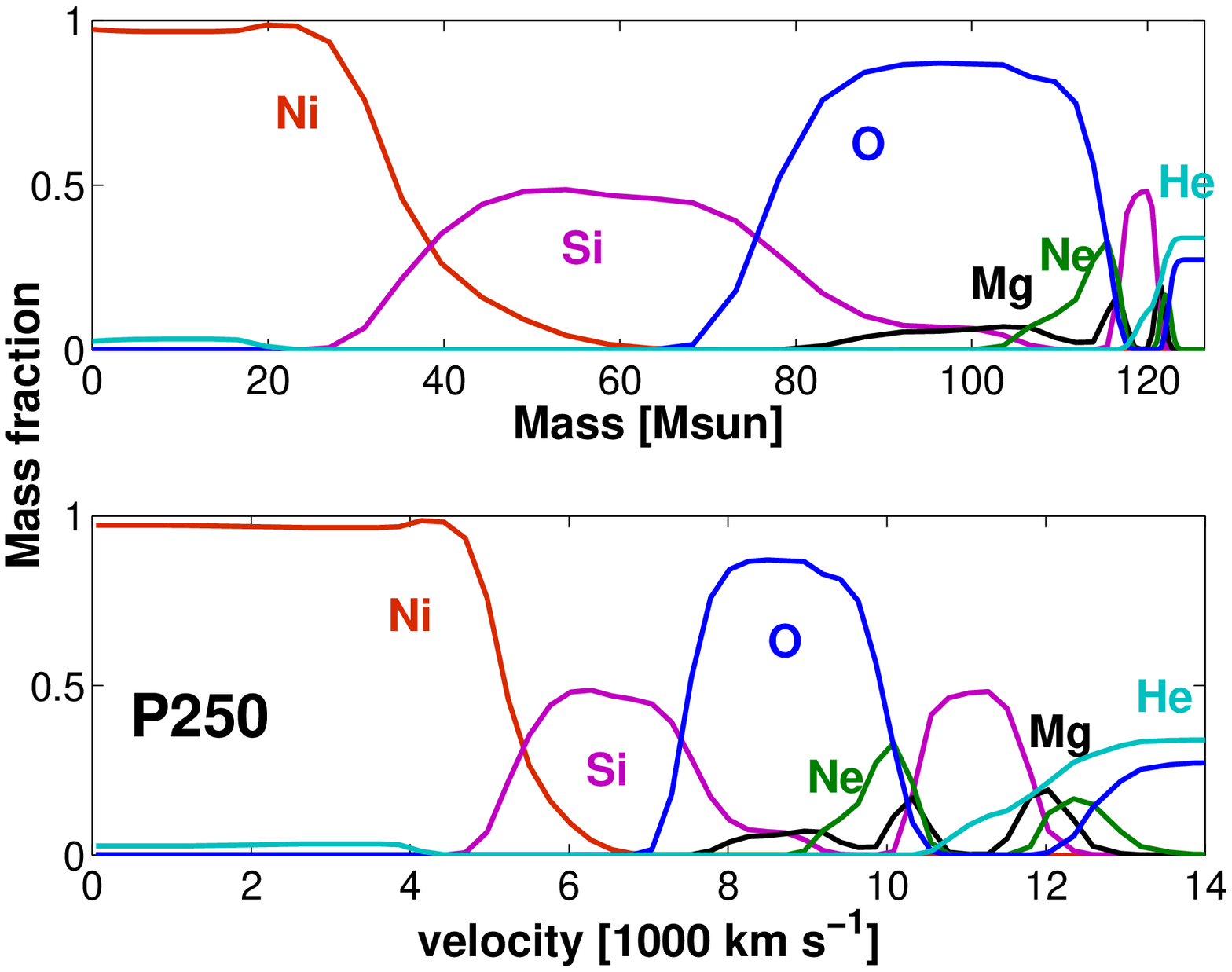}
\caption[Chemical structure of the P200 and P250 models]
{Chemical structure of the 200~\Msun{} (P200, left) and 250~\Msun{} (P250,right)
non-rotating PISN models simulated with \texttt{FLASH}
after the pair-instability explosion and when all nuclear burning is
over.  {\emph{Top}}: versus mass coordinate.  {\emph{Bottom}}: versus
radial velocity of the ejecta (in 1000~km\,s$^{\,-1}$).}
\label{figure:chemieP200P250}
\end{figure*}

In Fig.~\ref{figure:chemieP200P250}, we present the chemical structure of the models
as they were mapped into the \texttt{STELLA} code for calculating further
hydrodynamical and radiative evolution.  We
plot the most influential and abundant species -- helium, oxygen, neon, magnesium, silicon
and nickel-56.  The effects of the shock initiated nucleosynthesis can be
seen in the dips at the oxygen-helium interfaces where shock heating has
triggered $\alpha$--captures on carbon and oxygen producing silicon.  
Mass fractions at the surface of the model P250 are roughly: carbon
-- 0.39, helium -- 0.34, and oxygen -- 0.27, -- resulting from
convective helium shell burning.  The envelope above this layer was
completely lost due to the stellar wind during earlier evolutionary phases
\citep{2013MNRAS.433.1114Y}.  
We list the properties of our progenitor models and the
explosion results in Table~\ref{table:models}.

\subsection[Post-explosion dynamics with \texttt{STELLA}]{Post-explosion dynamics with \texttt{STELLA}}
\label{subsect:stella}

To simulate the supernova ejecta evolution and the light curves we used
the one--dimensional radiation hydrodynamics code
\texttt{STELLA} \citep[][]{2006A&A...453..229B,2014A&A...565A..70K}.  The
PISN models are mapped into \texttt{STELLA} before the shock
reaches the surface of the progenitor, i.e. just before shock breakout.
While mapping into \texttt{STELLA}, 
P200 and P250 models were divided in to 194 and 116 zones, respectively.

\texttt{STELLA} solves the radiative-transfer equations in the intensity
momentum approximation in each frequency bin.  We use 100 frequency groups in the current
study.  These are enough groups to produce spectral energy distribution, but are not
sufficient to produce spectra.  The opacity is computed based on about
153,441 spectral lines from \citet{1995all..book.....K} and
\citet{1996ADNDT..64....1V}.  The expansion opacity formalism from
\citet{1993ApJ...412..731E} is used for line opacity taking the effect of high velocity gradients
into account.  Opacity also
includes photoionization, free-free absorption, and electron scattering.
Local thermodynamic equilibrium (LTE) is assumed in the plasma, which allows the use of
the Boltzmann-Saha distribution for ionization and level populations.  
\texttt{STELLA} does not include a nuclear network except radioactive decay of
nickel-56 to cobalt-56, and to iron-56.  The code uses 16 species for calculating the overall
opacity.  These are: H, He, C, N, O, Ne, Na, Mg, Al, Si, S, Ar, Ca, 
a sum of stable Fe and radioactive $^{56}$Co, stable Ni, and radioactive $^{56}$Ni.  
Energy from nickel and cobalt radioactive decay is deposited 
into positrons and gamma-photons and is
treated in a one-group transport approximation according to \citet{1995ApJ...446..766S}. 

\texttt{STELLA} solves the conservation equations for mass, momentum, and total energy
in the Lagrangian co-moving grid.  
The artificial viscosity consists of the standard
von~Neumann artificial viscous pressure used for stabilizing solution \citep{1950vonNeumann}
and a so-called ``cold artificial viscosity'' used to smear shocks
\citep{1998ApJ...496..454B,2013PhDTMoriya}.  
Therefore, \texttt{STELLA} allows one to properly compute the propagation of the
shock along the ejecta and the shock-breakout event.   The coupled 
equations of radiation hydrodynamics (system of ordinary differential equations) are solved through an
implicit high-order predictor-corrector procedure based on the
methods of \citet{1971nivp.book.....G} and \citet{1972Brayton} \citep[see details
in][]{1996AstL...22...39B,Stabrowski1997}.  The required
accuracy is set at the level of $10^{\,-3}-10^{\,-4}$, whereas the actual accuracy
is better than 1\%.

\texttt{STELLA} was successfully applied to normal and peculiar SNe~Ia
\citep{2000AstL...26...67S,2006A&A...453..229B,2007PASP..119..360P}, SNe~IIP
\citep{2005AstL...31..429B,2016ApJ...821..124T}, and SNe~IIpec
\citep[SN~1987A, SN~1993J,][]{1998ApJ...496..454B,2000ApJ...532.1132B}, SNe~IIL
\citep{1993A&A...273..106B,2016MNRAS.455..423M}.  
Since \texttt{STELLA} is a hydrodynamics code, it is widely used for simulations
of interacting supernovae, in which normal supernova ejecta collide with a
shell or dense circumstellar environment or a wind
\citep{2011MNRAS.415..199M,2015AstL...41...95B,2015arXiv151000834S}.  
\citet{2014A&A...565A..70K} use
\texttt{STELLA} for simulating post-explosion radiation and hydrodynamical 
evolution of low-mass and high-mass hydrogen-rich PISNe.

In some of our simulations, the outermost layers of the supernova ejecta
reach very high velocities.  
In these cases, we truncated a small fraction of the outer layer,
to ensure stability of the \texttt{STELLA} simulations.  This  causes a slightly
weaker luminosity (since $L \sim R^2$) at the so-called ``plateau'' phase before re-brightening, 
but does not affect the main nickel-powered maximum, 
because the layer removed is almost massless and does not carry much kinetic energy.

\section[Results]{Results}
\label{sect:results}

\begin{figure*}
\centering
\includegraphics[width=0.5\textwidth]{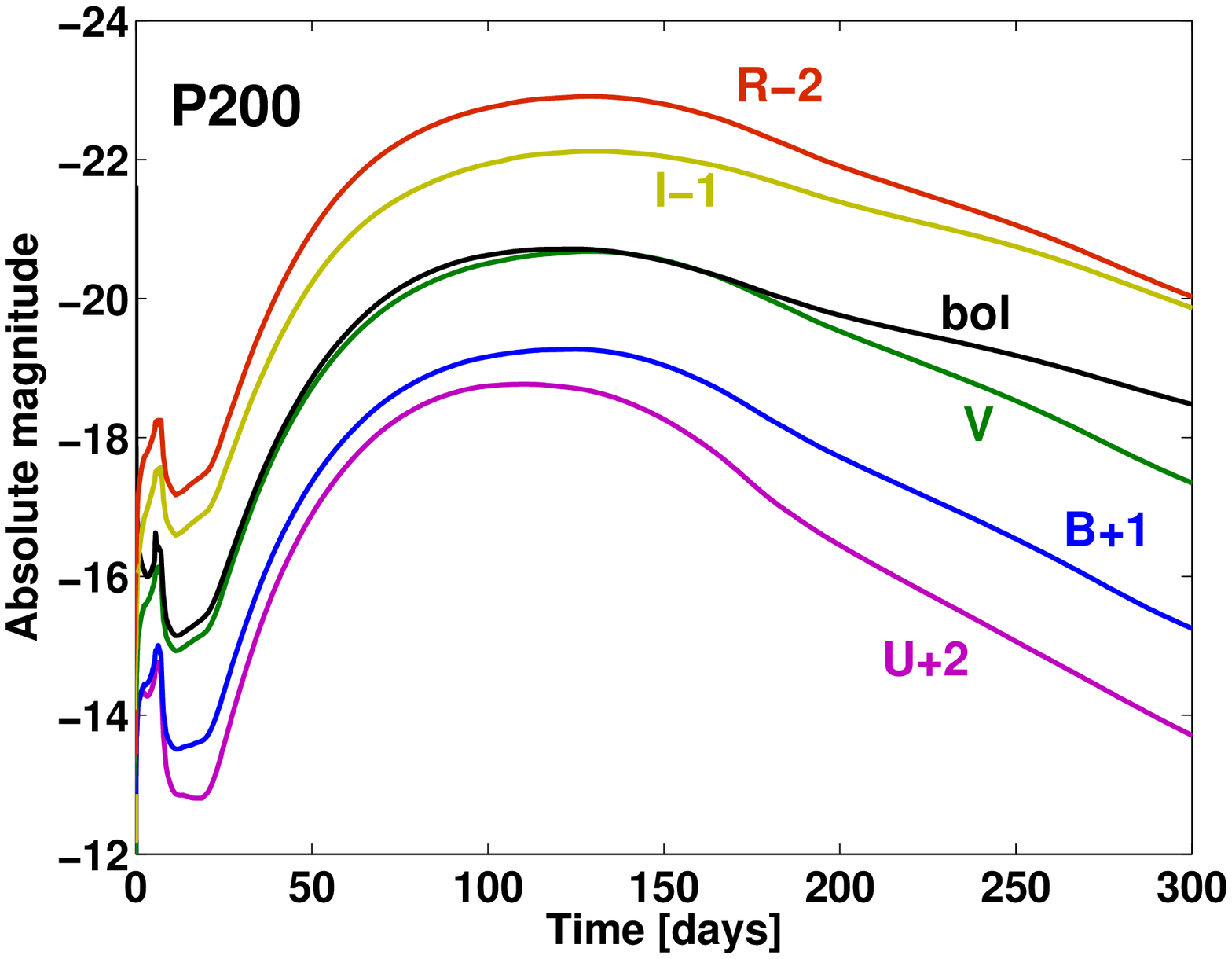}~~
\includegraphics[width=0.5\textwidth]{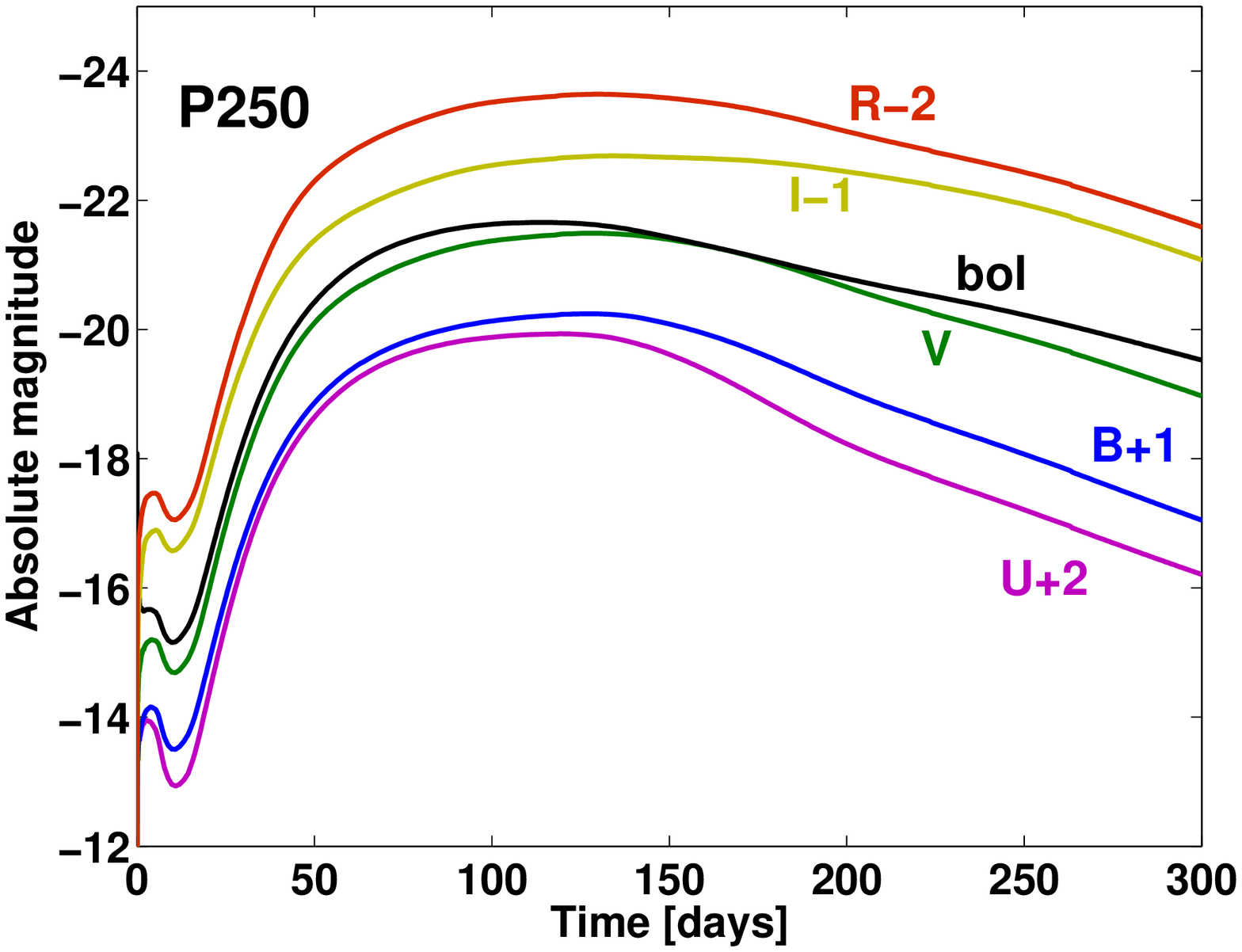}
\caption[P200 and P250 bolometric and {\emph{UBVRI}} broad band light curves]
{200~\Msun{} ({\emph{left}}: P200) and 250~\Msun{} ({\emph{right}}: P25) 
PISN bolometric (black) and {\emph{UBVRI}} broad band light curves.}
\label{figure:P200P250}
\end{figure*}

We show the bolometric and broad-band light curves for our main models
P200 and P250 in Fig.~\ref{figure:P200P250}.  P200 and P250 reach a maximum bolometric
luminosity of $6\times10^{\,43}$~erg\,s$^{\,-1}$ and
$1.4\times10^{\,44}$~erg\,s$^{\,-1}$, respectively.  
All figures start with time `0' which
corresponds to the time at the beginning of the \texttt{STELLA} simulations.  

We mapped the \texttt{FLASH} P200 and P250 outputs into \texttt{STELLA},
when the shock propagates through the outer layer.  
The shock reaches the surface during time $\approx R/v_\mathrm{sound}$, i.e.
almost immediately after mapping.  
According to \citet{2016ApJ...821..124T}, the duration of the shock breakout
event mostly depends on radius of the progenitor.  In the case of our compact
models, the shock breakout lasts about 9~minutes for P200 and 4~s for P250. 
Hence, in Fig.~\ref{figure:P200P250}, the light curves begin with the shock
breakout which remains unresolved on the plots because of relatively shorter
time-scale.

\begin{figure}
\centering
\includegraphics[width=0.5\textwidth]{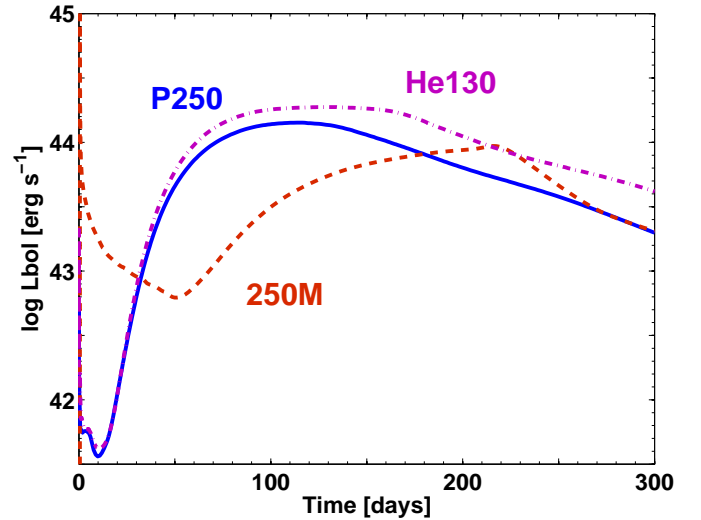}
\caption[Bolometric light curves for the P250, helium He130 and hydrogen-rich 250M PISN models]
{Bolometric light curves for the model P250 (solid), helium He130 (dash-dotted)
and the hydrogen-rich PISN model 250M \citep[dashed,][]{2014A&A...565A..70K}.  All
light curves are calculated with \texttt{STELLA}.}
\label{figure:250MP250}
\end{figure}

One of the most noticeable features of the present results is
the short rise time for the given PISN light curves.  The re-brightening
phase lasts about 100~days for both models, which is noticeably shorter
than for previously published light curves.  
For instance, all models presented in \citet{2013MNRAS.428.3227D} rise
to maximum during about 150--200~days, and models presented in \citet{2011ApJ...734..102K} rise
during 150--400~days dependent on the type of progenitor. 
In Fig.~\ref{figure:250MP250}, 
we include the long-rising curve for the hydrogen-rich model
250M \citep{2007A&A...475L..19L,2014A&A...565A..70K} and helium model He130
\citep[][]{2011ApJ...734..102K} together with the P250 curve for illustration.  The long
rise time disfavours PISNe as a possible scenario  to explain SLSNe.  
However, our new light curves of P200 and P250 PISNe evolve faster
than hydrogen-rich PISNe and might be more relevant to at least some of the observed SLSNe.  
We explain the faster evolution of the P200 and P250 light curves by the very 
distinct distribution of hydrogen, helium and
nickel-56 in the P200 and P250 models as explained below.  
It is well-known that hydrogen is the most influential element
supporting the electron-scattering opacity
and governing the location of the photosphere.  If hydrogen is absent, 
helium dominates the electron-scattering opacity.   
At the same time, the nickel-56 distribution also strongly impacts the light curve
appearance, especially during rise.  
The model 250M retains 58~\Msun{} of hydrogen-helium in the envelope which significantly
impedes inward motion of the photosphere and delays re-brightening to the nickel powered
maximum for 200~d, while the P250 model has
only 2~\Msun{} of helium in its atmosphere.  The surface abundances in P250 are
dominated by carbon and oxygen, with a mass fraction of 0.34 of helium.  
Radioactive material is distributed in up to 
half of the P250 ejecta by mass coordinate, and up to 30\% of the
250M ejecta.  The combination of a small helium layer and closeness of
radioactive material to the surface of the progenitor leads to a fast evolving
light curve for the P250 model compared to the slowly evolving 250M model. 
Therefore, the 100-day rise time makes the new PISN models, presented in the 
current study, as good candidates for explanation of some SLSNe. 

The chemical structure of the model P250 resembles that of the model
He130, although $^{56}$Ni mass is higher in He130 than in P250, and surface helium mass
fraction differs considerably.  
The higher $^{56}$Ni mass leads to
a broader peak, and the higher surface helium
abundance in He130 causes a longer rise for the He130 light curve.  
Therefore, the P250 and He130 light curves differ.  
In Fig.~\ref{figure:250MP250}, all light curves of 250M, P250 and He130
models are simulated with \texttt{STELLA}.  The He130 light curve published
earlier was computed with \texttt{SEDONA} \citep[][]{2011ApJ...734..102K}.  
The uncertainty of the results due to the different radiation codes will
be discussed in Section~\ref{sect:code2code}.

The photosphere in our new models is located deep in the oxygen layer (close to
the bottom of oxygen shell), therefore, P200 and P250 explosions appear as hydrogen and
helium-free at maximum light, i.e. as Type~I supernovae.   
We discuss the applicability of the P200 and P250 models to SLSN PTF12dam in the
next section.
\section[Comparison to SLSN PTF12dam]{Comparison to SLSN PTF12dam}
\label{sect:compare}

\begin{figure}
\centering
\includegraphics[width=0.5\textwidth]{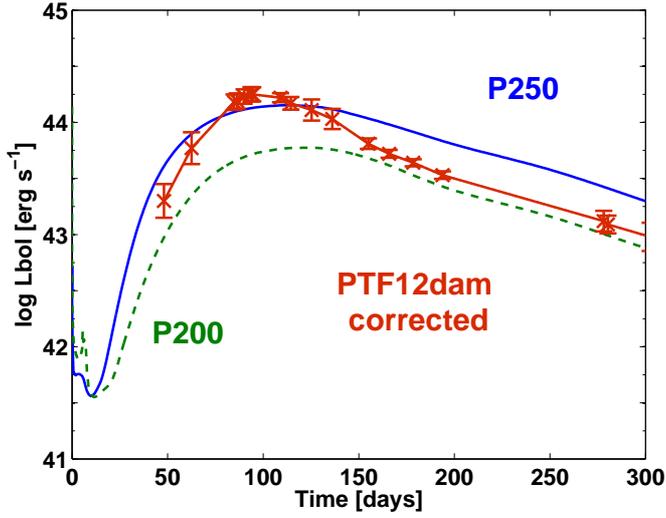}
\caption[Bolometric light curves for the P200 and P250 PISN models versus SLSN~PTF12dam]
{Bolometric light curves for the P200 and P250 PISN models and SLSN~PTF12dam.}
\label{figure:P250PTF12dam_L}
\end{figure}

\begin{figure}
\centering
\includegraphics[width=0.5\textwidth]{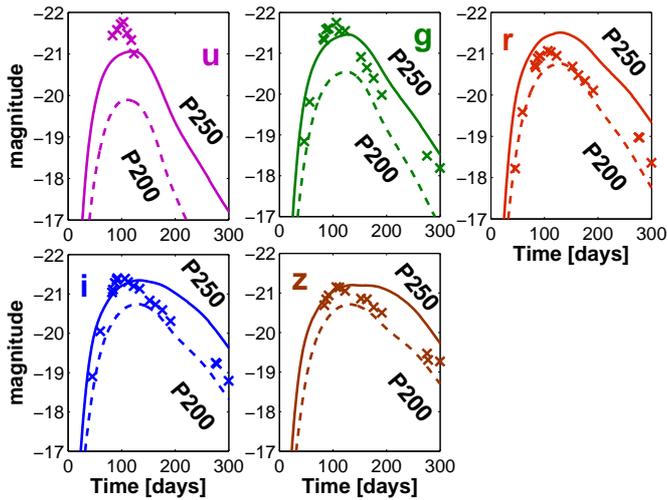}
\caption[P250 vs. SLSN~PTF12dam in $ugriz$-bands]
{P200 (dashed) and P250 (solid) vs. SLSN~PTF12dam (crosses) in $ugriz$-bands.}
\label{figure:P250PTF12dam_bands}
\end{figure}

Inspired by the short rise time of the P200 and P250 light curves and considerably
high luminosity, we decided to put our models into the context of SLSNe.  We
choose SLSN~PTF12dam as it is one of the
well-observed recent SLSNe \citep{2013Natur.502..346N,2015MNRAS.452.1567C}.
In Fig.~\ref{figure:P250PTF12dam_L}, we show the comparison of our models
with the bolometric light curve of PTF12dam.  
The observed data are
shifted by 100~days, allowing the observed peak luminosity to approximately
coincide with the maximum of the P250 synthetic light curve.  
The figure demonstrates that the shape of the bolometric synthetic light curves
resembles the behaviour of the observed light curve of PTF12dam around maximum epoch.  
Fig.~\ref{figure:P250PTF12dam_bands} shows
the synthetic curves of P200 and P250 in $ugriz$-bands and observed
absolute $ugriz$-magnitudes of PTF12dam.

\begin{figure}
\centering
\includegraphics[width=0.5\textwidth]{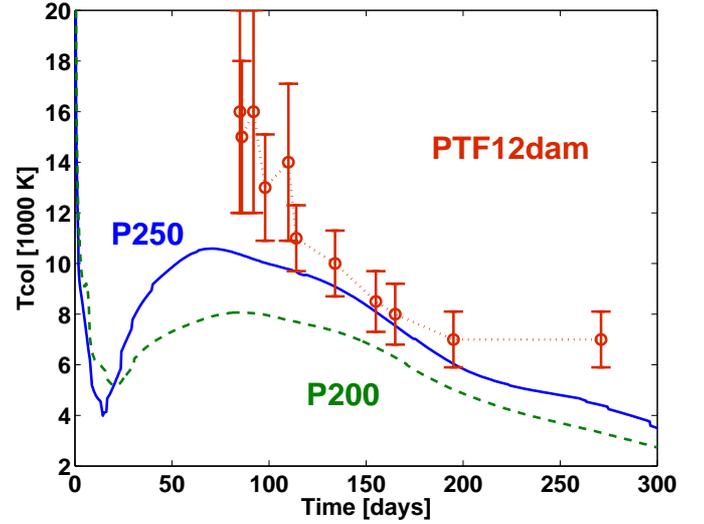}
\caption[Colour temperature for the P200 and P250 PISN models versus SLSN~PTF12dam]
{Colour temperature the P200 and P250 PISN models and that of SLSN~PTF12dam.}
\label{figure:P200P250Tcol}
\end{figure}

\begin{figure}
\centering
\includegraphics[width=0.5\textwidth]{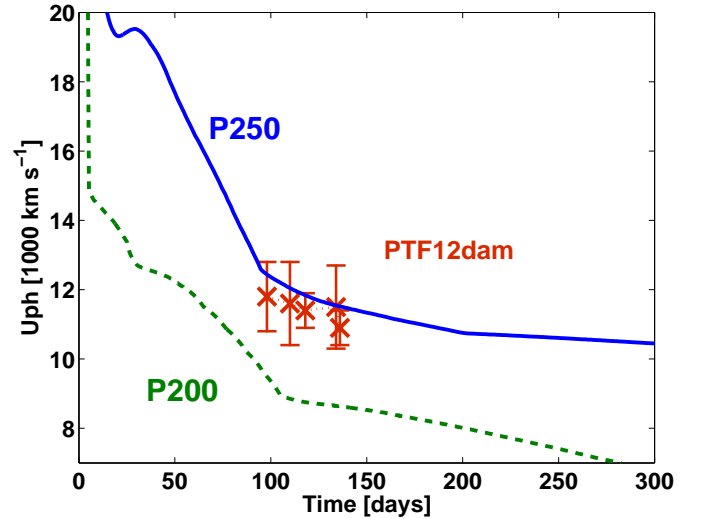}
\caption[P200 and P250 PISN photospheric velocity evolution vs. SLSN~PTF12dam]
{Photospheric velocity evolution for the P200 and P250 PISN models versus PTF12dam data.}
\label{figure:P200P250uph}
\end{figure}

Figs.~\ref{figure:P200P250Tcol} and \ref{figure:P200P250uph} show colour temperature
and photospheric velocity evolution for the P200 and P250 models versus
those of PTF12dam \citep{2013Natur.502..346N}.  
We estimate the colour temperature based on the least-square method using the
spectral range from 1 to 50,000~\AA{}.  The colour temperature reaches 
8,100\,K in the P200 model and 11,000\,K in the P250 model at peak
luminosity, which is higher compared to previously
published PISN models.  
The photospheric velocity is the radial velocity of the layer where the photosphere is located.  
The photospheric velocity is 9,000\,km\,s$^{\,-1}$ (P200) and 
12,000\,km\,s$^{\,-1}$ (P250) at peak luminosity, respectively.


P200 and P250 models reproduce parts of the PTF12dam data.  
In particular, P250 matches the earlier bolometric light curve to some
degree, the peak luminosity,
the colour temperature of P250 is close to the data points during
100~days after the bolometric peak, while photospheric velocity in P250 ejecta fully
matches the observed velocity.  P200 model better matches the late part of
the light curve.  Some features, however, are not well explained by the
models.  The colour temperature near the peak of the light curve is not
matched by the models.  The broad band light curves also do not match very
well, although this is difficult for any model to explain.  
We emphasize that our
models are computed self-consistently and without fine tuning for
PTF12dam.  We conclude that the PISN scenario is still viable for
PTF12dam\footnote{The newest PTF12dam data published very recently by
\citet{2016arXiv160908145V} show that
the bolometric light curve is slightly broader, and the colour temperature is
noticeably lower, reaching only 11,700~K at maximum luminosity, which may favour our P250 model.}.



\section[Modelling uncertainties and comparison with other radiation codes]
{Comparison between different numerical approaches}
\label{sect:code2code}

As we show in Section~\ref{sect:results}, our new PISN models exhibit
relatively short rise to the main nickel-powered maximum compared to previously
published PISN light curves.  In order to assess the robustness of these findings, we
confront \texttt{Stella} calculations to the results obtained with different
numerical approaches used to solve the radiative transport problem in supernova
ejecta.  Here, we mainly focus on \texttt{SEDONA} which has been extensively
used to predict PISN observables.  All technical details of the different
methods used in the following analysis are deferred to the
Appendix~\ref{sect:append1}.



\subsection[The input for comparison: Helium PISN He130]
{The reference model: Helium PISN He130}
\label{subsect:He130}

All calculations performed in this comparative analysis are based on the
130~\Msun{} helium PISN model He130 since it resembles our P250 PISN model
fairly closely and since it is a well-accepted model in the PISN context.  This
model has been simulated with the \texttt{KEPLER} stellar evolution code
\citep{1978ApJ...225.1021W,2002RvMP...74.1015W} from the helium main sequence,
i.e.\ as a pure helium star without any wind mass-loss, through the
pair-instability phase.  It retains a shallow outer shell of 1.65~\Msun{} helium
and produces 40~\Msun{} of radioactive nickel-56.

\subsection{Simple Test Calculation}
\label{subsect:codecmp}

\begin{figure}
\centering
\includegraphics[width=0.5\textwidth]{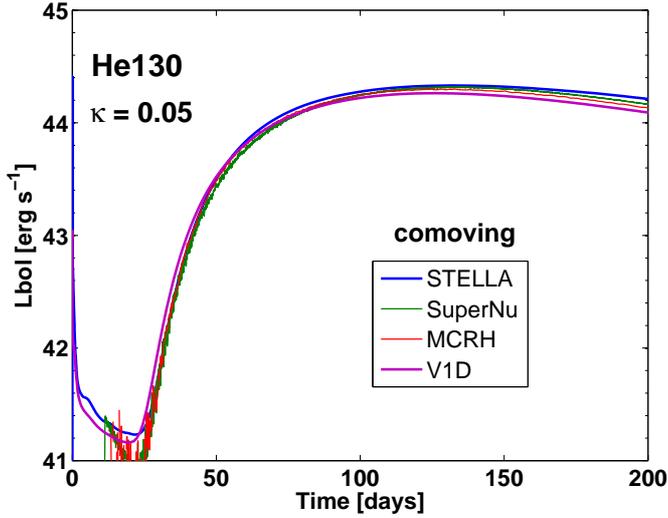}
\caption[\texttt{SuperNu}, \texttt{STELLA}, \texttt{V1D} and \texttt{MCRH} simulations
with the constant opacity]{\texttt{SuperNu}, \texttt{STELLA}, \texttt{V1D} and
\texttt{MCRH} simulations of the He130 model with the constant opacity,
$\kappa=0.05$ in the comoving frame.}  
\label{figure:Kappa}
\end{figure}

As a preparation, we avoid complications induced by different assumptions about
ionization and excitation and by the details of the opacity treatments by
construction a simple test problem based on the He130 model.  In particular, we
assume a constant, frequency-independent specific interaction cross section
($\kappa=0.05$~cm$^{\,2}$\,g$^{\,-1}$) and run simulations with
\texttt{STELLA}, \texttt{V1D}, and two Monte-Carlo codes \texttt{MCRH} and
\texttt{SuperNu} (see description of the codes in Appendix~\ref{sect:append1}).  
The results are presented in Figure~\ref{figure:Kappa} showing an excellent
agreement between the bolometric light curves computed with all the different
methods.  Thus, when adopting the same physical assumptions, \texttt{STELLA}
performs as well as other radiative transfer methods.

\subsection[\texttt{STELLA} versus \texttt{SEDONA}]
{\texttt{STELLA} versus \texttt{SEDONA}}
\label{subsect:sedona}

\begin{figure}
\centering
\includegraphics[width=0.5\textwidth]{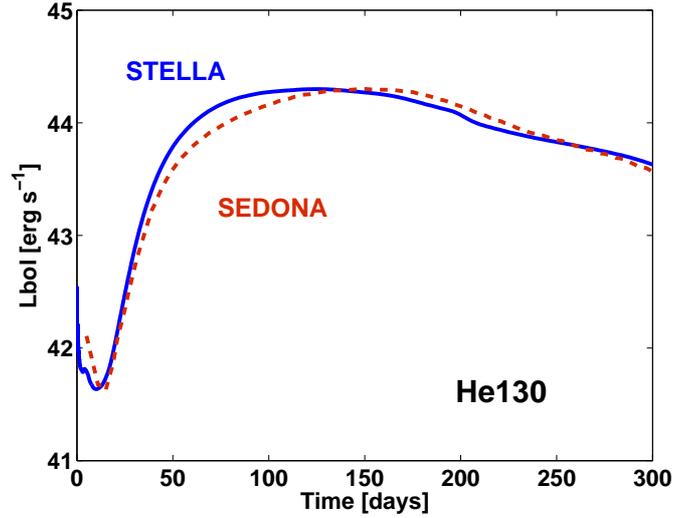}
\caption[Comparison between STELLA and SEDONA]{He130 bolometric light curves
with \texttt{STELLA} and \texttt{SEDONA} codes.  See discussion in
the text.}
\label{figure:Kasen}
\end{figure}

\begin{figure}
\centering
\includegraphics[width=0.5\textwidth]{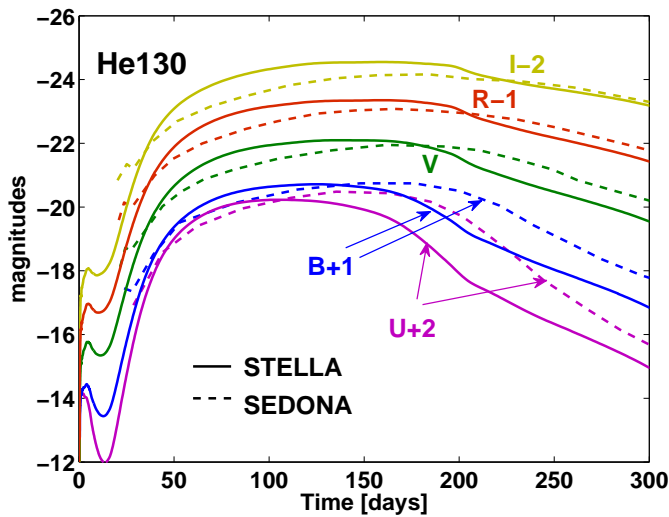}
\caption[Comparison between STELLA and SEDONA]{He130 in broad bands 
with \texttt{STELLA} (solid) and \texttt{SEDONA} (dashed) codes.  
The light curves are, from bottom to top, {\emph{UBVRI}}.  
\emph{U}, \emph{B}, \emph{V}, \emph{R}, and \emph{I},
magnitudes are plotted with a shift of +2, +1, 0, -1, -2~magnitudes, respectively.}
\label{figure:Kasenbands}
\end{figure}

Having completed the first comparison under idealised conditions, we turn to
calculations under more realistic conditions.  In particular, we compute the
evolution of the PISN model He130 with \texttt{STELLA} starting at 100~s after
the pair-instability explosion \citep{2002ApJ...567..532H}.  To avoid problems
associated with relativistic effects, we truncate the initial \texttt{KEPLER}
profile at about 10\% of speed of light.  However, velocity exceeds this limit
after the shock breaks out and reaches $5\times10^{\,9}$\,cm\,s$^{\,-1}$.  The
obtained bolometric light curve is compared to the published \texttt{SEDONA}
results in Fig.~\ref{figure:Kasen}.  The overall width and shape of the two
light curves are in good agreement.  If compared in detail, however, the
bolometric light curve of He130 model seems to rise again faster when computed
with \texttt{STELLA}.  The difference amounts to approximately 50~days.  When
comparing the two calculations in different broad bands, the discrepancies
become a bit more noticeable as seen in Fig.~\ref{figure:Kasenbands}.  This is
not too surprising given the differences the detailed radiative transfer
treatments (ionization and excitation prescriptions, opacity treatments, atomic
data etc.).

In a series of additional calculations, we investigate this difference in the
early light curve evolution in more detail.  In particular, we examine whether
deviations from homology are causing this and to which extent details in the
opacity treatment play a role in this context.

\subsection[Influence of Deviation from Homology - Nickel Bubble Effect]
{Influence of Deviation from Homology - Nickel Bubble Effect}
\label{subsect:homo}

Unlike the SEDONA version used to calculate the published light curves of the
PISN model He130\footnote{Recently, \citet{2015ApJS..217....9R} have successfully
developed a one-dimensional radiation hydrodynamics version of \texttt{SEDONA}},
\texttt{STELLA} solves the full radiation hydrodynamical problem and is thus
able to track deviation from homologous expansion.  The radiation released in
the radioactive decay will exert a pressure on the surrounding ejecta material
as it diffuses out and will thus inflate nickel-rich regions.  We now
investigate the influence of this radiation hydrodynamical effect on the PISN
light curve, in particular on the rise time.

\begin{figure*}
\centering
\includegraphics[width=0.5\textwidth]{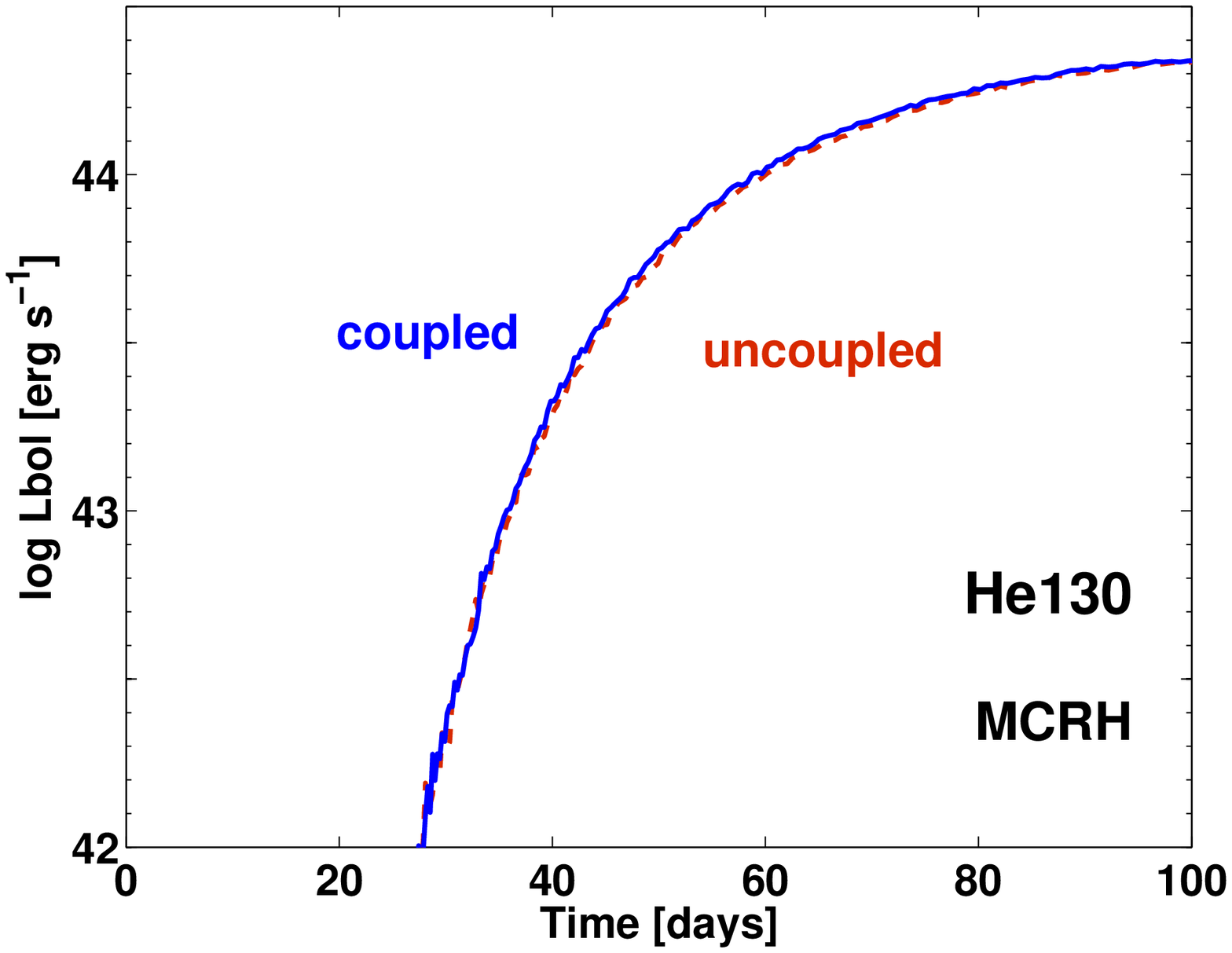}~~
\includegraphics[width=0.5\textwidth]{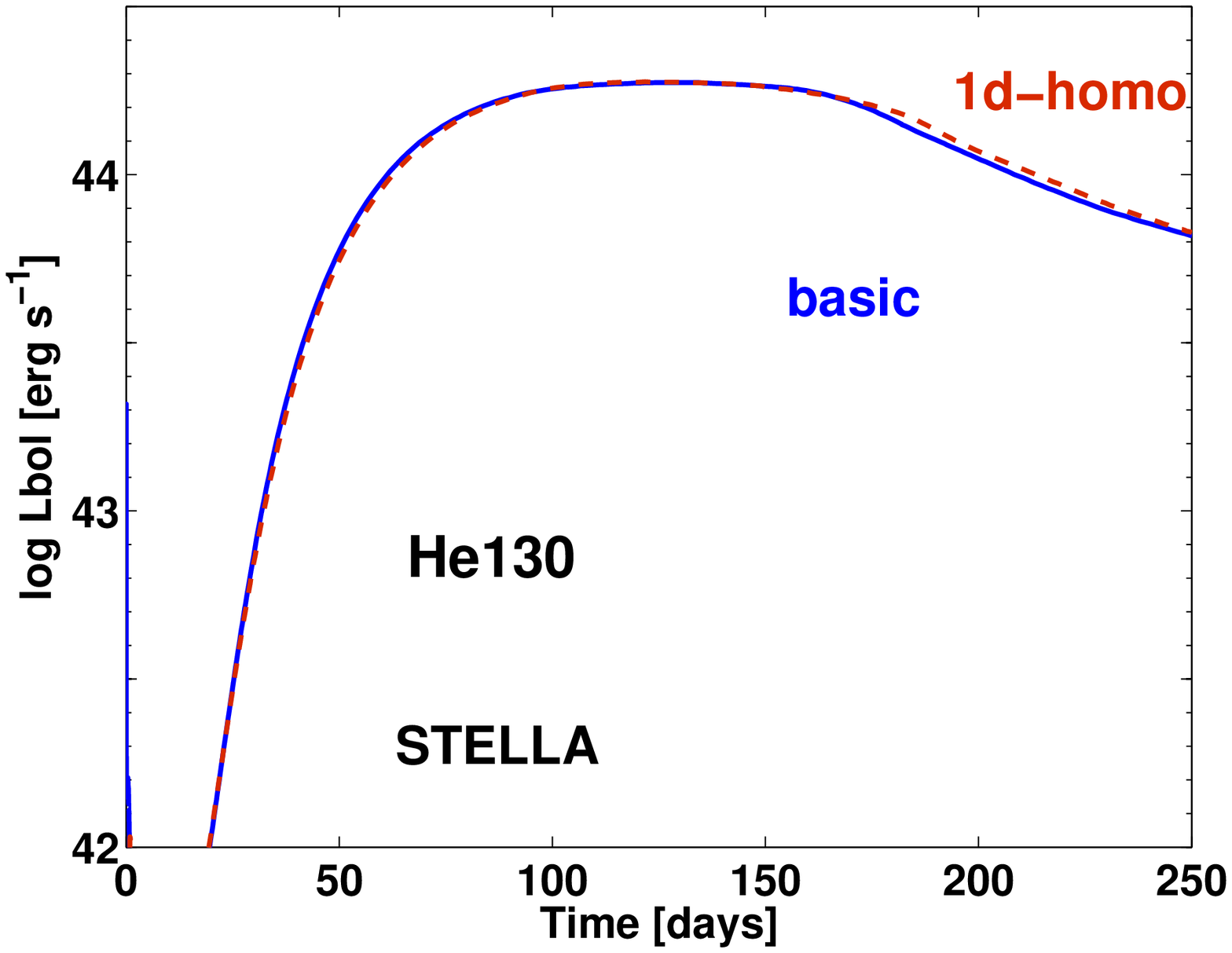}
\caption[Monte-Carlo radiation hydrodynamics simulation with and without
coupling]{He130 bolometric light curves simulated with and without
coupling.  {\emph{Left:}} Simulations undertaken with the Monte-Carlo radiation
hydrodynamics code \texttt{MCRH} \citep{2012MNRAS.425.1430N}. 
{\emph{Right:}} Simulations undertaken with \texttt{STELLA}.  
Label `1d-homo' indicates run in which homologous expansion starts at day~1.}
\label{figure:homo}
\end{figure*}

For this purpose, we recalculate the He130 model with the Monte Carlo based
radiation-hydrodynamics code \texttt{MCRH} and 
determine the influence of deviations from homology
on the emergent light curve analogously to
\citet{2012MNRAS.425.1430N}, where this effect has been explored in the SNe~Ia
context.  In particular, bolometric light curves are calculated once assuming
pure homologous expansion and switching the radiation hydrodynamical
coupling off and a second time with the coupling taken into account.  For these
\texttt{MCRH} calculations, a constant, frequency-independent specific
interaction cross section was adopted ($\kappa =
0.1\,\mathrm{cm^{\,2}\,g^{\,-1}}$).  More technical details about MCRH and the
simulations are provided in the Appendix~\ref{sect:append1}.  The left panel of
Figure~\ref{figure:homo} shows a comparison and demonstrates that deviation
from homology seem to have insignificant consequences on the emergent PISN
light curve.  This finding is confirmed by an additional test calculation
performed with \texttt{STELLA}.  Here, the hydrodynamical coupling has been
artificially suppressed after day~1{}.  As seen in the right panel of
Figure~\ref{figure:homo}, the resulting light curve is almost identical to the
one obtained in the full \texttt{STELLA} simulation.

\begin{figure}
\centering
\includegraphics[width=0.5\textwidth]{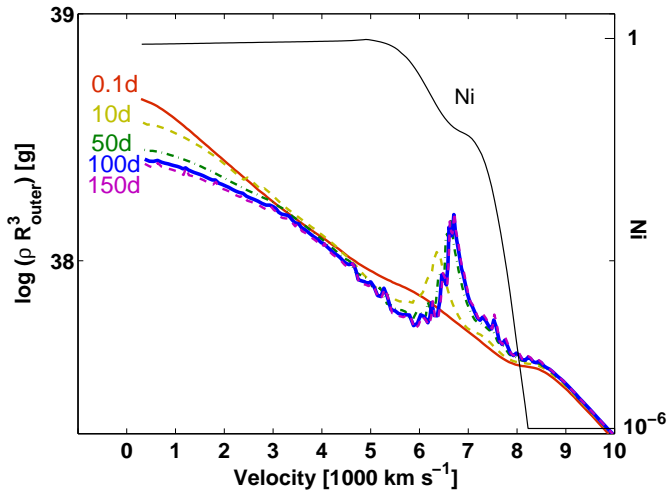}
\caption[Nickel-bubble effect]{Scaled density--velocity evolution of the inner
ejecta of the He130 model
simulated with \texttt{STELLA} according to the nickel-bubble effect.  
Density is scaled to a factor of $R_{\mathrm{outer}}^{\,3}$.  In addition the nickel
distribution (not to scale) is shown as a thin line.  See discussion in the text.}
\label{figure:Nibubble}
\end{figure}

We emphasize, that even though radiation hydrodynamical effects do not seem to
play a role for calculation of bolometric light curves, the density structure
is significantly modified by the dynamical effect of the radiation generated in
the radioactive decay.  The radiation field supplies an additional contribution
to the pressure and inflates nickel-rich regions.  The ``nickel bubble effect''
\citep[well-explained in][]{2006A&A...453..229B,2007ApJ...662..487W} develops
during the first 100~days after the explosion in the He130 model as illustrated
in Fig.~\ref{figure:Nibubble}.  In the inner regions of the He130 model, the
density is decreased relative to homologous expansion by a factor of 2 and the
velocity is boosted by about 25\%.  This dilutes the central nickel bubble and
increases its radius by up to 40\%.  Above this central region, at about
7,000~km\,s$^{\,-1}$, a narrow shell with enhanced density is generated,
containing mostly silicon, sulphur and oxygen.  This phenomenon might impact the
spectrum formation \citep[see discussion in][]{2016arXiv160802994J}.

\subsection[Influence of Opacity]{Influence of Opacity}
\label{subsect:lists}

\begin{figure*}
\centering
\includegraphics[width=0.5\textwidth]{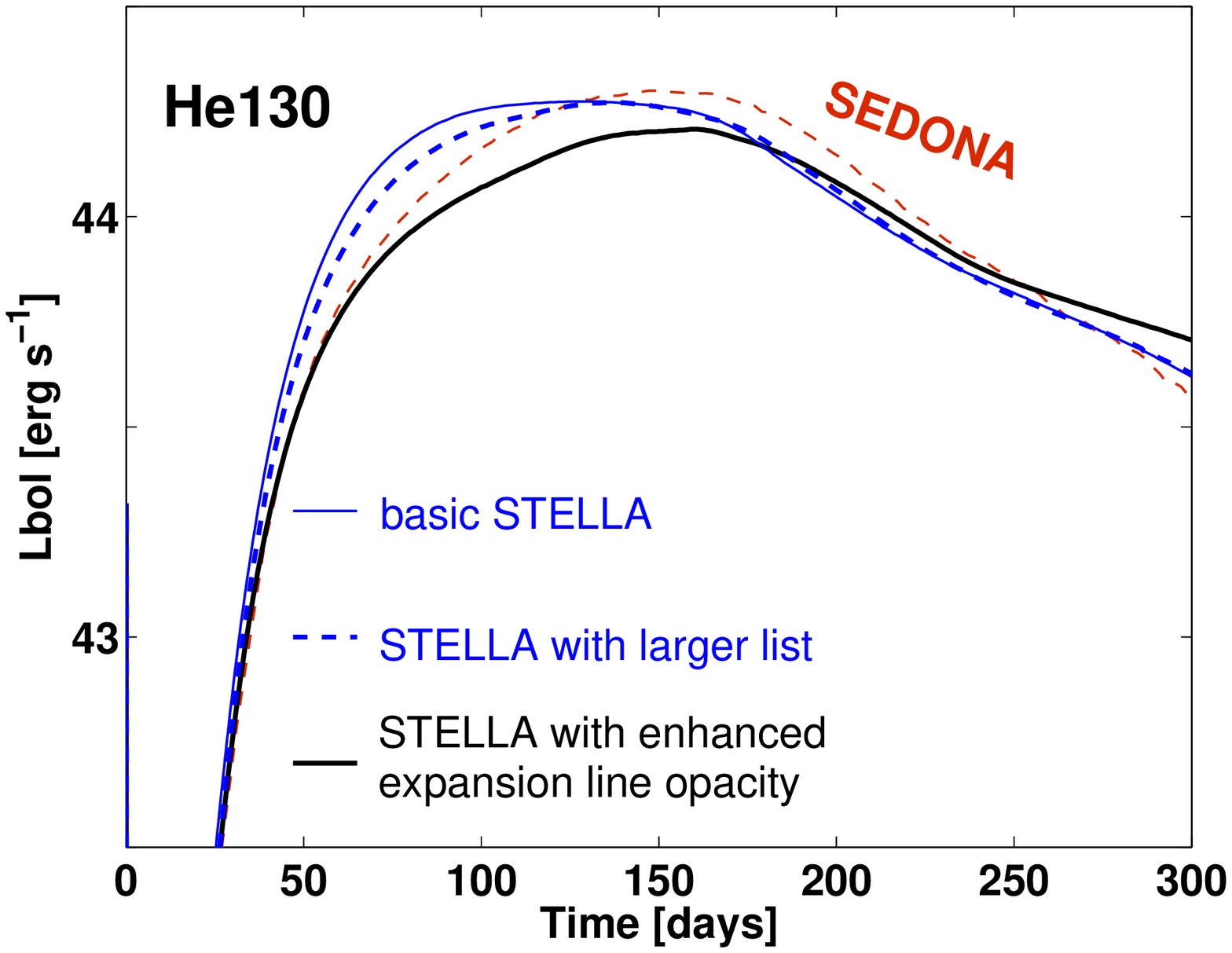}~~
\includegraphics[width=0.5\textwidth]{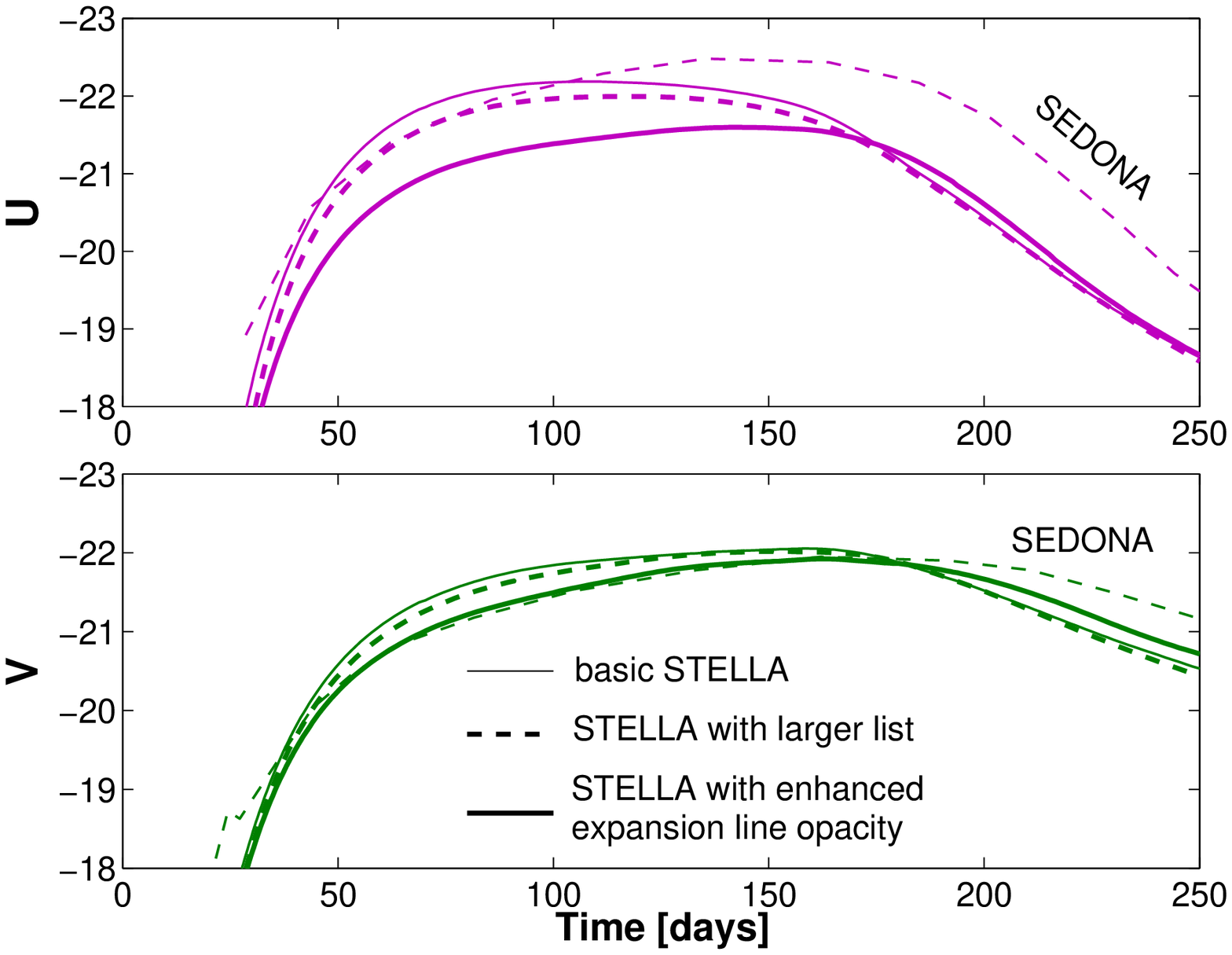}
\caption[Comparison between \texttt{SEDONA} and \texttt{STELLA} light curves with different
line lists]{Bolometric (left), and {\emph{U}} and {\emph{V}} broad band (right) light curve for
He130 model computed with \texttt{SEDONA} (thin dashed),
\texttt{STELLA} with the basic \texttt{STELLA} list of lines (thin solid), 
\texttt{STELLA} with the extended list of lines (thick dashed), and
\texttt{STELLA} with the enhanced basic line-opacity (thick solid).}
\label{figure:difflist}
\end{figure*}


Naturally, the radiation simulations strongly depend on the underlying opacity.
Therefore, we carried out additional simulations with \texttt{STELLA}, in which
we implemented a list containing 317,700 transitions from the 
\citet{1995all..book.....K} database.  As illustrated in Figure~\ref{figure:difflist}, the
\texttt{STELLA} bolometric light curve computed with the extended line-list
tends to become more similar in shape to the \texttt{SEDONA} bolometric light
curve.  In the right panel of Figure~\ref{figure:difflist}, we show light
curves in the {\emph{U}} and {\emph{V}} broad bands for illustration.  There
are some differences between the light curves in the {\emph{U}} band calculated
with \texttt{STELLA} using the basic line-list and the extended line-list,
while light the curves in the {\emph{B}}, {\emph{V}}, {\emph{R}}, {\emph{I}}
bands have minor changes.  

Even though the \texttt{STELLA} light curves obtained with the extended line-list resemble the
published \texttt{SEDONA} results more closely, there are still considerable
differences.  Considering the opacity treatment, the inclusion of millions of
weak line transitions in an expansion opacity formalism on top of the several
hundred thousand lines which are treated in detail in \texttt{SEDONA} may play
a role here.  To illustrate this, we also carried out \texttt{STELLA}
simulations with the basic line-list, in which the stronger line opacity is
mimicked by fixing velocity gradient on day~10.  The resulting bolometric light
curve is shown as the black solid line in the left panel of
Fig.~\ref{figure:difflist}, and thick solid lines in the right panel.  The
increased opacity delays the maximum and makes the light curve shallower during
the re-brightening phase.

From these explorations, we conclude that the basic \texttt{STELLA} spectral
line list contains all strong lines which govern the supernova light curve
during the photospheric phase and provides quite reliable resulting bolometric
light curves and magnitudes in broad bands on the time-scale from shock
breakout to several hundreds days.  However, the detailed shape of the light
curve, from which diagnostics such as the rise time is derived, is sensitive to
the details of the opacity treatment, for example to the number of line
transitions taken into account.

\subsection{Summary of the Code Comparison Experiments}

Based on the test calculations presented in the section we conclude that:

\begin{itemize}
  \item when adopting the same physical assumptions, in particular when
    considering the idealised situation with a constant, frequency-independent
    specific interaction cross section, the \texttt{STELLA} bolometric light
    curve agrees very well  with those computed with comparable radiative
    transfer and radiation hydrodynamics codes;
  \item since \texttt{STELLA} solves the coupled evolution of hydrodynamics and
    radiative transfer, the so-called nickel-bubble effect is seen in the
    \texttt{STELLA} calculations.  This process changes the ejecta structure
    noticeably but has no significant effect on the bolometric light curve as
    various test calculations demonstrated;
  \item the direct comparison of \texttt{STELLA} and \texttt{SEDONA}
    calculations seems to point to systematic differences in the rising part of
    the light curve.  Our test calculations indicate that details of the opacity
    treatment seem to play an important role in this context.
\end{itemize}

\section[Conclusions]{Conclusions}
\label{sect:conclusions}

In the present study we computed the evolution, explosion and post-explosion
evolution and light curves for two non-rotating stellar models with initial
masses 200~\Msun{} (P200) and 250~\Msun{} (P250) at a metallicity $Z=0.001$.  
For that we consecutively used the stellar evolution code \texttt{GENEC}, the
hydrodynamics code \texttt{FLASH}, and the radiation hydrodynamics code
\texttt{STELLA}.  P200 and P250 lose their entire hydrogen-rich
envelope due to radiatively-driven winds.  
P200 and P250 retain only 9~\Msun{} and 2~\Msun{} of helium just before the
pair-instability explosion in their outer layers.  During the explosion, P200
and P250 produced 12~\Msun{} and 34~\Msun{} of radioactive nickel, thus
powering luminous supernovae.  P200 and P250 reach peak luminosities of
$6\times10^{\,43}$~erg\,s$^{\,-1}$ and $1.4\times10^{\,44}$~erg\,s$^{\,-1}$,
respectively.  The colour temperature is 8,100~K (P200) and 11,000~K (P250) at
the maximum light.  As the photosphere resides at the bottom of oxygen
shell at the peak luminosity, the P200 and P250 explosions appear as hydrogen
and helium-free (Type~I) supernovae.

An important result of our study is the short rise time and fast evolution of
the light curves.  In particular, we find in our \texttt{STELLA} light curve
calculations that P200 and P250 rise to maximum in about one hundred days.  This
finding, that light curves of PISNe models which do not retain hydrogen at the
time of explosion evolve much faster than their hydrogen-rich siblings, is
compatible with previous studies \citep{2011ApJ...734..102K,2013MNRAS.428.3227D}.  
The short rise found in our calculations is a consequence of 
\begin{itemize}
\item the absence of hydrogen,
\item a relatively shallow helium layer, 
\item an extended nickel distribution.
\end{itemize}
Note that we do not apply any artificial mixing in our \texttt{FLASH} and
\texttt{STELLA} simulations of P200 and P250.  

We examine the short rise of the \texttt{STELLA} light curve, by
carrying out additional simulations of the helium He130 PISN model from
\citet{2002ApJ...567..532H}.  The
nickel-bubble effect has an impact on the density and velocity profiles and hydrodynamics
but a negligible effect on the light curve properties.  The treatment of
opacities has a noticeable impact on the light curve.  Artificially enforcing a
constant specific interaction cross section enables us to obtain very similar
light curves with four different codes (\texttt{STELLA}, \texttt{MCRH},
\texttt{SuperNu}, and \texttt{V1D}) for the He130 progenitor model.  Increasing
the number of lines in the line-list included in \texttt{STELLA} lengthens the
rise time but does not explain the full difference between \texttt{STELLA} and
\texttt{SEDONA}.  Nevertheless these calculations together with 
the artificially enhanced line opacities demonstrate that the opacity has 
the strongest effect
on the light curve shape around maximum.  Additionally, differences and
uncertainties in the progenitor structure also affect the peak of the light
curve and thus indirectly the rise time.  Possibly the slope during the rise of
the light curve is a more robust feature.  Despite these uncertainties, we
confirm that hydrogen-free PISN light curves evolve faster than those of hydrogen-rich
PISNe, possibly fast enough to explain SLSNe such as PTF12dam.   

We compare P200 and P250 models to the well-observed SLSN PTF12dam.  From our
analysis, P200 and P250 models reproduce parts of the PTF12dam data.  P250
matches the earlier bolometric light curve to some degree, the peak luminosity,
the colour temperature of P250 is close to the data points during 100~days
after the bolometric peak, while photospheric velocity in P250 ejecta fully
matches the observed velocity.  P200 model better matches the late part of the
light curve.  
To conclude, pair-instability supernova scenario can still be
a reasonable candidate for explaining observables of PTF12dam.  
The very massive (above 60~\Msun{}) stellar origin of this event was proposed by
\citet[][]{2015MNRAS.451L..65T} and \citet{2016arXiv160802994J}, 
as the supernova exploded in the
star-forming region of a fairly low metallicity dwarf galaxy.  
Other models proposed to explain PTF12dam are the
magnetar-powered models 
\citep{2013Natur.502..346N,2013MNRAS.432.3228K,2014MNRAS.437..703M} and interaction-driven models
\citep{2013ApJ...773...76C,2015AstL...41...95B,2015MNRAS.452.1567C,2016Tolstov}.  
The next test for our models will be to compute spectra for the
photospheric phase or/and for the nebular phase.  We will present the
nebular spectrum simulations in the forthcoming paper
(Mazzali \& Kozyreva, in preparation).

\section*{Acknowledgements}


The \texttt{STELLA} simulations were particularly carried out on the DIRAC Complexity system,
operated by the University of
Leicester IT Services, which forms part of the STFC DiRAC HPC Facility
(\url{www.dirac.ac.uk}).  This equipment is funded by BIS National
E-Infrastructure capital grant ST/K000373/1 and STFC DiRAC Operations grant
ST/K0003259/1.  DiRAC is part of the National E-Infrastructure.  
AK and RH acknowledge support from EU-FP7-ERC-2012-St Grant~306901.  
MG and CF acknowledge support from the Department of Energy through an
Early Career Award (DOE grant No.~SC0010263).  
The work of SB on development of the {\texttt{STELLA}} code is supported by a
grant from the Russian Science Foundation (project number 14-12-00203).
Work at LANL (WPE, RTW) was done under the auspices of the National Nuclear Security
Administration of the U.S. Department of Energy at Los Alamos National
Laboratory under Contract No. DE-AC52-06NA25396.  All LANL calculations 
were performed on Institutional Computing resources.  
UMN is supported by the Transregional Collaborative Research Centre TRR 33 ``The Dark Universe''
of the German Research Foundation (DFG).  
DRvR is supported in part at the University of Chicago by the National
Science Foundation under grants AST--0909132, PHY--0822648 (JINA, Joint
Institute for Nuclear Astrophysics), and PHY--1430152 (JINA-CEE, Joint
Institute for Nuclear Astrophysics).  
AH is supported by an ARC Future Fellowship (FT120100363).  
AT is supported by the World Premier International Research Center Initiative (WPI Initiative, MEXT, Japan).  
EC is supported by Enrico Fermi Fellow.  
AK is grateful to Andrea~Cristini for proofreading the
manuscript, Stuart Sim, Markus Kromer, Stefan Taubenberger, Claes Fransson, Daniel Whalen, 
Luc Dessart and Daniel Kasen for fruitful discussions and useful suggestions.

\addcontentsline{toc}{section}{Acknowledgements}

\bibliographystyle{mnras}

\begin{thebibliography}{}
\makeatletter
\relax
\def\mn@urlcharsother{\let\do\@makeother \do\$\do\&\do\#\do\^\do\_\do\%\do\~}
\def\mn@doi{\begingroup\mn@urlcharsother \@ifnextchar [ {\mn@doi@}
  {\mn@doi@[]}}
\def\mn@doi@[#1]#2{\def\@tempa{#1}\ifx\@tempa\@empty \href
  {http://dx.doi.org/#2} {doi:#2}\else \href {http://dx.doi.org/#2} {#1}\fi
  \endgroup}
\def\mn@eprint#1#2{\mn@eprint@#1:#2::\@nil}
\def\mn@eprint@arXiv#1{\href {http://arxiv.org/abs/#1} {{\tt arXiv:#1}}}
\def\mn@eprint@dblp#1{\href {http://dblp.uni-trier.de/rec/bibtex/#1.xml}
  {dblp:#1}}
\def\mn@eprint@#1:#2:#3:#4\@nil{\def\@tempa {#1}\def\@tempb {#2}\def\@tempc
  {#3}\ifx \@tempc \@empty \let \@tempc \@tempb \let \@tempb \@tempa \fi \ifx
  \@tempb \@empty \def\@tempb {arXiv}\fi \@ifundefined
  {mn@eprint@\@tempb}{\@tempb:\@tempc}{\expandafter \expandafter \csname
  mn@eprint@\@tempb\endcsname \expandafter{\@tempc}}}

\bibitem[\protect\citeauthoryear{Abdikamalov, Burrows, Ott, Loffler, O'Connor,
  Dolence  \& Schnetter}{Abdikamalov et~al.}{2012}]{abdikamalov2012}
Abdikamalov E.,  Burrows A.,  Ott C.~D.,  Loffler F.,  O'Connor E.,  Dolence
  J.~C.,   Schnetter E.,  2012, ApJ, 755, 111

\bibitem[\protect\citeauthoryear{{Angulo} et~al.,}{{Angulo}
  et~al.}{1999}]{1999NuPhA.656....3A}
{Angulo} C.,  et~al., 1999, \mn@doi [Nuclear Physics A]
  {10.1016/S0375-9474(99)00030-5}, \href
  {http://adsabs.harvard.edu/abs/1999NuPhA.656....3A} {656, 3}

\bibitem[\protect\citeauthoryear{{Asplund}, {Grevesse}  \& {Sauval}}{{Asplund}
  et~al.}{2005}]{APS05}
{Asplund} M.,  {Grevesse} N.,   {Sauval} A.~J.,  2005, in {Barnes} III T.~G.,
  {Bash} F.~N.,  eds,  Astronomical Society of the Pacific Conference Series
  Vol. 336, Cosmic Abundances as Records of Stellar Evolution and
  Nucleosynthesis. p.~25

\bibitem[\protect\citeauthoryear{{Baklanov}, {Blinnikov}  \&
  {Pavlyuk}}{{Baklanov} et~al.}{2005}]{2005AstL...31..429B}
{Baklanov} P.~V.,  {Blinnikov} S.~I.,   {Pavlyuk} N.~N.,  2005, \mn@doi
  [Astronomy Letters] {10.1134/1.1958107}, \href
  {http://adsabs.harvard.edu/abs/2005AstL...31..429B} {31, 429}

\bibitem[\protect\citeauthoryear{{Baklanov}, {Sorokina}  \&
  {Blinnikov}}{{Baklanov} et~al.}{2015}]{2015AstL...41...95B}
{Baklanov} P.~V.,  {Sorokina} E.~I.,   {Blinnikov} S.~I.,  2015, \mn@doi
  [Astronomy Letters] {10.1134/S1063773715040027}, \href
  {http://adsabs.harvard.edu/abs/2015AstL...41...95B} {41, 95}

\bibitem[\protect\citeauthoryear{{Baraffe}, {Heger}  \& {Woosley}}{{Baraffe}
  et~al.}{2001}]{2001ApJ...550..890B}
{Baraffe} I.,  {Heger} A.,   {Woosley} S.~E.,  2001, \mn@doi [\apj]
  {10.1086/319808}, \href {http://adsabs.harvard.edu/abs/2001ApJ...550..890B}
  {550, 890}

\bibitem[\protect\citeauthoryear{{Barkat}, {Rakavy}  \& {Sack}}{{Barkat}
  et~al.}{1967}]{1967PhRvL..18..379B}
{Barkat} Z.,  {Rakavy} G.,   {Sack} N.,  1967, \mn@doi [Physical Review
  Letters] {10.1103/PhysRevLett.18.379}, \href
  {http://adsabs.harvard.edu/abs/1967PhRvL..18..379B} {18, 379}

\bibitem[\protect\citeauthoryear{{Bisnovatyi-Kogan} \&
  {Kazhdan}}{{Bisnovatyi-Kogan} \& {Kazhdan}}{1967}]{1967SvA....10..604B}
{Bisnovatyi-Kogan} G.~S.,  {Kazhdan} Y.~M.,  1967, \sovast, \href
  {http://adsabs.harvard.edu/abs/1967SvA....10..604B} {10, 604}

\bibitem[\protect\citeauthoryear{{Blinnikov} \& {Bartunov}}{{Blinnikov} \&
  {Bartunov}}{1993}]{1993A&A...273..106B}
{Blinnikov} S.~I.,  {Bartunov} O.~S.,  1993, \aap, \href
  {http://adsabs.harvard.edu/abs/1993A%26A...273..106B} {273, 106}

\bibitem[\protect\citeauthoryear{{Blinnikov} \& {Panov}}{{Blinnikov} \&
  {Panov}}{1996}]{1996AstL...22...39B}
{Blinnikov} S.~I.,  {Panov} I.~V.,  1996, Astronomy Letters, \href
  {http://adsabs.harvard.edu/abs/1996AstL...22...39B} {22, 39}

\bibitem[\protect\citeauthoryear{{Blinnikov}, {Dunina-Barkovskaya}  \&
  {Nadyozhin}}{{Blinnikov} et~al.}{1996}]{1996ApJS..106..171B}
{Blinnikov} S.~I.,  {Dunina-Barkovskaya} N.~V.,   {Nadyozhin} D.~K.,  1996,
  \mn@doi [\apjs] {10.1086/192334}, \href
  {http://adsabs.harvard.edu/abs/1996ApJS..106..171B} {106, 171}

\bibitem[\protect\citeauthoryear{{Blinnikov}, {Eastman}, {Bartunov},
  {Popolitov}  \& {Woosley}}{{Blinnikov} et~al.}{1998}]{1998ApJ...496..454B}
{Blinnikov} S.~I.,  {Eastman} R.,  {Bartunov} O.~S.,  {Popolitov} V.~A.,
  {Woosley} S.~E.,  1998, \mn@doi [\apj] {10.1086/305375}, \href
  {http://adsabs.harvard.edu/abs/1998ApJ...496..454B} {496, 454}

\bibitem[\protect\citeauthoryear{{Blinnikov}, {Lundqvist}, {Bartunov}, {Nomoto}
   \& {Iwamoto}}{{Blinnikov} et~al.}{2000}]{2000ApJ...532.1132B}
{Blinnikov} S.,  {Lundqvist} P.,  {Bartunov} O.,  {Nomoto} K.,   {Iwamoto} K.,
  2000, \mn@doi [\apj] {10.1086/308588}, \href
  {http://adsabs.harvard.edu/abs/2000ApJ...532.1132B} {532, 1132}

\bibitem[\protect\citeauthoryear{{Blinnikov}, {R{\"o}pke}, {Sorokina},
  {Gieseler}, {Reinecke}, {Travaglio}, {Hillebrandt}  \&
  {Stritzinger}}{{Blinnikov} et~al.}{2006}]{2006A&A...453..229B}
{Blinnikov} S.~I.,  {R{\"o}pke} F.~K.,  {Sorokina} E.~I.,  {Gieseler} M.,
  {Reinecke} M.,  {Travaglio} C.,  {Hillebrandt} W.,   {Stritzinger} M.,  2006,
  \mn@doi [\aap] {10.1051/0004-6361:20054594}, \href
  {http://adsabs.harvard.edu/abs/2006A%26A...453..229B} {453, 229}

\bibitem[\protect\citeauthoryear{{Bond}, {Arnett}  \& {Carr}}{{Bond}
  et~al.}{1982}]{1982sscr.conf..303B}
{Bond} J.~R.,  {Arnett} W.~D.,   {Carr} B.~J.,  1982, in {Rees} M.~J.,
  {Stoneham} R.~J.,  eds, NATO ASIC Proc. 90: Supernovae: A Survey of Current
  Research. pp 303--311

\bibitem[\protect\citeauthoryear{{Brayton}, {Gustavson}  \&
  {Hatchel}}{{Brayton} et~al.}{1972}]{1972Brayton}
{Brayton} R.~K.,  {Gustavson} F.~G.,   {Hatchel} G.~D.,  1972, in Proceedings
  of the IEEE.

\bibitem[\protect\citeauthoryear{{Carr}, {Bond}  \& {Arnett}}{{Carr}
  et~al.}{1984}]{1984ApJ...277..445C}
{Carr} B.~J.,  {Bond} J.~R.,   {Arnett} W.~D.,  1984, \mn@doi [\apj]
  {10.1086/161713}, \href {http://adsabs.harvard.edu/abs/1984ApJ...277..445C}
  {277, 445}

\bibitem[\protect\citeauthoryear{{Chatzopoulos}, {Wheeler}, {Vinko}, {Horvath}
  \& {Nagy}}{{Chatzopoulos} et~al.}{2013a}]{2013ApJ...773...76C}
{Chatzopoulos} E.,  {Wheeler} J.~C.,  {Vinko} J.,  {Horvath} Z.~L.,   {Nagy}
  A.,  2013a, \mn@doi [\apj] {10.1088/0004-637X/773/1/76}, \href
  {http://adsabs.harvard.edu/abs/2013ApJ...773...76C} {773, 76}

\bibitem[\protect\citeauthoryear{{Chatzopoulos}, {Wheeler}  \&
  {Couch}}{{Chatzopoulos} et~al.}{2013b}]{2013ApJ...776..129C}
{Chatzopoulos} E.,  {Wheeler} J.~C.,   {Couch} S.~M.,  2013b, \mn@doi [\apj]
  {10.1088/0004-637X/776/2/129}, \href
  {http://adsabs.harvard.edu/abs/2013ApJ...776..129C} {776, 129}

\bibitem[\protect\citeauthoryear{{Chatzopoulos}, {van Rossum}, {Craig},
  {Whalen}, {Smidt}  \& {Wiggins}}{{Chatzopoulos}
  et~al.}{2015}]{2015ApJ...799...18C}
{Chatzopoulos} E.,  {van Rossum} D.~R.,  {Craig} W.~J.,  {Whalen} D.~J.,
  {Smidt} J.,   {Wiggins} B.,  2015, \mn@doi [\apj]
  {10.1088/0004-637X/799/1/18}, \href
  {http://adsabs.harvard.edu/abs/2015ApJ...799...18C} {799, 18}

\bibitem[\protect\citeauthoryear{{Chen} et~al.,}{{Chen}
  et~al.}{2015}]{2015MNRAS.452.1567C}
{Chen} T.-W.,  et~al., 2015, \mn@doi [\mnras] {10.1093/mnras/stv1360}, \href
  {http://adsabs.harvard.edu/abs/2015MNRAS.452.1567C} {452, 1567}

\bibitem[\protect\citeauthoryear{{Colella} \& {Woodward}}{{Colella} \&
  {Woodward}}{1984}]{1984JCoPh..54..174C}
{Colella} P.,  {Woodward} P.~R.,  1984, \mn@doi [Journal of Computational
  Physics] {10.1016/0021-9991(84)90143-8}, \href
  {http://adsabs.harvard.edu/abs/1984JCoPh..54..174C} {54, 174}

\bibitem[\protect\citeauthoryear{{Crowther}, {Schnurr}, {Hirschi}, {Yusof},
  {Parker}, {Goodwin}  \& {Kassim}}{{Crowther}
  et~al.}{2010}]{2010MNRAS.408..731C}
{Crowther} P.~A.,  {Schnurr} O.,  {Hirschi} R.,  {Yusof} N.,  {Parker} R.~J.,
  {Goodwin} S.~P.,   {Kassim} H.~A.,  2010, \mn@doi [\mnras]
  {10.1111/j.1365-2966.2010.17167.x}, \href
  {http://adsabs.harvard.edu/abs/2010MNRAS.408..731C} {408, 731}

\bibitem[\protect\citeauthoryear{{Cunha}, {Hubeny}  \& {Lanz}}{{Cunha}
  et~al.}{2006}]{Cunha06}
{Cunha} K.,  {Hubeny} I.,   {Lanz} T.,  2006, \mn@doi [\apjl] {10.1086/507301},
  \href {http://adsabs.harvard.edu/abs/2006ApJ...647L.143C} {647, L143}

\bibitem[\protect\citeauthoryear{Densmore, Urbatsch, Evans  \& Buksas}{Densmore
  et~al.}{2007}]{densmore2007}
Densmore J.~D.,  Urbatsch T.~J.,  Evans T.~M.,   Buksas M.~W.,  2007, J.
  Comput. Phys., 222, 485

\bibitem[\protect\citeauthoryear{Densmore, Thompson  \& Urbatsch}{Densmore
  et~al.}{2012}]{densmore2012}
Densmore J.~D.,  Thompson K.~G.,   Urbatsch T.~J.,  2012, J. Comput. Phys.,
  231, 6925

\bibitem[\protect\citeauthoryear{{Dessart} \& {Hillier}}{{Dessart} \&
  {Hillier}}{2010}]{2010MNRAS.405.2141D}
{Dessart} L.,  {Hillier} D.~J.,  2010, \mn@doi [\mnras]
  {10.1111/j.1365-2966.2010.16611.x}, \href
  {http://adsabs.harvard.edu/abs/2010MNRAS.405.2141D} {405, 2141}

\bibitem[\protect\citeauthoryear{{Dessart}, {Livne}  \& {Waldman}}{{Dessart}
  et~al.}{2010}]{2010MNRAS.405.2113D}
{Dessart} L.,  {Livne} E.,   {Waldman} R.,  2010, \mn@doi [\mnras]
  {10.1111/j.1365-2966.2010.16626.x}, \href
  {http://adsabs.harvard.edu/abs/2010MNRAS.405.2113D} {405, 2113}

\bibitem[\protect\citeauthoryear{{Dessart}, {Hillier}, {Waldman}, {Livne}  \&
  {Blondin}}{{Dessart} et~al.}{2012}]{2012MNRAS.426L..76D}
{Dessart} L.,  {Hillier} D.~J.,  {Waldman} R.,  {Livne} E.,   {Blondin} S.,
  2012, \mn@doi [\mnras] {10.1111/j.1745-3933.2012.01329.x}, \href
  {http://adsabs.harvard.edu/abs/2012MNRAS.426L..76D} {426, L76}

\bibitem[\protect\citeauthoryear{{Dessart}, {Waldman}, {Livne}, {Hillier}  \&
  {Blondin}}{{Dessart} et~al.}{2013}]{2013MNRAS.428.3227D}
{Dessart} L.,  {Waldman} R.,  {Livne} E.,  {Hillier} D.~J.,   {Blondin} S.,
  2013, \mn@doi [\mnras] {10.1093/mnras/sts269}, \href
  {http://adsabs.harvard.edu/abs/2013MNRAS.428.3227D} {428, 3227}

\bibitem[\protect\citeauthoryear{{Dessart}, {Audit}  \& {Hillier}}{{Dessart}
  et~al.}{2015}]{2015MNRAS.449.4304D}
{Dessart} L.,  {Audit} E.,   {Hillier} D.~J.,  2015, \mn@doi [\mnras]
  {10.1093/mnras/stv609}, \href
  {http://adsabs.harvard.edu/abs/2015MNRAS.449.4304D} {449, 4304}

\bibitem[\protect\citeauthoryear{{Dubey}, {Reid}, {Weide}, {Antypas},
  {Ganapathy}, {Riley}, {Sheeler}  \& {Siegal}}{{Dubey}
  et~al.}{2009}]{2009arXiv0903.4875D}
{Dubey} A.,  {Reid} L.~B.,  {Weide} K.,  {Antypas} K.,  {Ganapathy} M.~K.,
  {Riley} K.,  {Sheeler} D.,   {Siegal} A.,  2009, preprint (\mn@eprint {arXiv}
  {0903.4875})

\bibitem[\protect\citeauthoryear{{Eastman} \& {Pinto}}{{Eastman} \&
  {Pinto}}{1993}]{1993ApJ...412..731E}
{Eastman} R.~G.,  {Pinto} P.~A.,  1993, \mn@doi [\apj] {10.1086/172957}, \href
  {http://adsabs.harvard.edu/abs/1993ApJ...412..731E} {412, 731}

\bibitem[\protect\citeauthoryear{{Ekstr{\"o}m} et~al.,}{{Ekstr{\"o}m}
  et~al.}{2012}]{2012A&A...537A.146E}
{Ekstr{\"o}m} S.,  et~al., 2012, \mn@doi [\aap] {10.1051/0004-6361/201117751},
  \href {http://adsabs.harvard.edu/abs/2012A%26A...537A.146E} {537, A146}

\bibitem[\protect\citeauthoryear{{Ferguson}, {Alexander}, {Allard}, {Barman},
  {Bodnarik}, {Hauschildt}, {Heffner-Wong}  \& {Tamanai}}{{Ferguson}
  et~al.}{2005}]{AF05}
{Ferguson} J.~W.,  {Alexander} D.~R.,  {Allard} F.,  {Barman} T.,  {Bodnarik}
  J.~G.,  {Hauschildt} P.~H.,  {Heffner-Wong} A.,   {Tamanai} A.,  2005,
  \mn@doi [\apj] {10.1086/428642}, \href
  {http://adsabs.harvard.edu/abs/2005ApJ...623..585F} {623, 585}

\bibitem[\protect\citeauthoryear{Fleck \& Cummings}{Fleck \&
  Cummings}{1971}]{fleck1971}
Fleck Jr. J.~A.,  Cummings J.~D.,  1971, J. Comput. Phys., 8, 313

\bibitem[\protect\citeauthoryear{{Fraley}}{{Fraley}}{1968}]{1968Ap&SS...2...96%
F}
{Fraley} G.~S.,  1968, \mn@doi [\apss] {10.1007/BF00651498}, \href
  {http://adsabs.harvard.edu/abs/1968Ap%26SS...2...96F} {2, 96}

\bibitem[\protect\citeauthoryear{{Fryxell}, {M{\"u}ller}  \&
  {Arnett}}{{Fryxell} et~al.}{1989}]{1989nuas.conf..100F}
{Fryxell} B.,  {M{\"u}ller} E.,   {Arnett} D.,  1989, in {Hillebrandt} W.,
  {M{\"u}ller} E.,  eds, Nuclear Astrophysics.

\bibitem[\protect\citeauthoryear{{Fryxell} et~al.,}{{Fryxell}
  et~al.}{2000}]{2000ApJS..131..273F}
{Fryxell} B.,  et~al., 2000, \mn@doi [\apjs] {10.1086/317361}, \href
  {http://adsabs.harvard.edu/abs/2000ApJS..131..273F} {131, 273}

\bibitem[\protect\citeauthoryear{{Gal-Yam}}{{Gal-Yam}}{2012}]{2012Sci...337..9%
27G}
{Gal-Yam} A.,  2012, \mn@doi [Science] {10.1126/science.1203601}, \href
  {http://adsabs.harvard.edu/abs/2012Sci...337..927G} {337, 927}

\bibitem[\protect\citeauthoryear{{Gal-Yam}, {Mazzali}, {Manulis}  \&
  {Bishop}}{{Gal-Yam} et~al.}{2013}]{2013PASP..125..749G}
{Gal-Yam} A.,  {Mazzali} P.~A.,  {Manulis} I.,   {Bishop} D.,  2013, \mn@doi
  [\pasp] {10.1086/671483}, \href
  {http://adsabs.harvard.edu/abs/2013PASP..125..749G} {125, 749}

\bibitem[\protect\citeauthoryear{{Gear}}{{Gear}}{1971}]{1971nivp.book.....G}
{Gear} C.~W.,  1971, {Numerical initial value problems in ordinary differential
  equations}

\bibitem[\protect\citeauthoryear{{Glatzel} \& {Kiriakidis}}{{Glatzel} \&
  {Kiriakidis}}{1993}]{1993MNRAS.263..375G}
{Glatzel} W.,  {Kiriakidis} M.,  1993, \mn@doi [\mnras]
  {10.1093/mnras/263.2.375}, \href
  {http://adsabs.harvard.edu/abs/1993MNRAS.263..375G} {263, 375}

\bibitem[\protect\citeauthoryear{{Gr{\"a}fener} \& {Hamann}}{{Gr{\"a}fener} \&
  {Hamann}}{2008}]{GH08}
{Gr{\"a}fener} G.,  {Hamann} W.-R.,  2008, \mn@doi [\aap]
  {10.1051/0004-6361:20066176}, \href
  {http://adsabs.harvard.edu/abs/2008A%26A...482..945G} {482, 945}

\bibitem[\protect\citeauthoryear{{Habibi}, {Stolte}  \& {Harfst}}{{Habibi}
  et~al.}{2014}]{2014A&A...566A...6H}
{Habibi} M.,  {Stolte} A.,   {Harfst} S.,  2014, \mn@doi [\aap]
  {10.1051/0004-6361/201323030}, \href
  {http://adsabs.harvard.edu/abs/2014A%26A...566A...6H} {566, A6}

\bibitem[\protect\citeauthoryear{{Heger} \& {Woosley}}{{Heger} \&
  {Woosley}}{2002}]{2002ApJ...567..532H}
{Heger} A.,  {Woosley} S.~E.,  2002, \mn@doi [\apj] {10.1086/338487}, \href
  {http://adsabs.harvard.edu/abs/2002ApJ...567..532H} {567, 532}

\bibitem[\protect\citeauthoryear{{Hirschi}}{{Hirschi}}{2007}]{2007A&A...461..5%
71H}
{Hirschi} R.,  2007, \mn@doi [\aap] {10.1051/0004-6361:20065356}, \href
  {http://adsabs.harvard.edu/abs/2007A%26A...461..571H} {461, 571}

\bibitem[\protect\citeauthoryear{{Hirschi}}{{Hirschi}}{2015}]{2015ASSL..412..1%
57H}
{Hirschi} R.,  2015, in {Vink} J.~S.,  ed.,  Astrophysics and Space Science
  Library Vol. 412, Very Massive Stars in the Local Universe. p.~157
  (\mn@eprint {arXiv} {1409.7053}), \mn@doi{10.1007/978-3-319-09596-7_6}

\bibitem[\protect\citeauthoryear{{Humphreys} \& {Davidson}}{{Humphreys} \&
  {Davidson}}{1994}]{1994PASP..106.1025H}
{Humphreys} R.~M.,  {Davidson} K.,  1994, \mn@doi [\pasp] {10.1086/133478},
  \href {http://adsabs.harvard.edu/abs/1994PASP..106.1025H} {106, 1025}

\bibitem[\protect\citeauthoryear{{Iglesias} \& {Rogers}}{{Iglesias} \&
  {Rogers}}{1996}]{1996ApJ...464..943I}
{Iglesias} C.~A.,  {Rogers} F.~J.,  1996, \mn@doi [\apj] {10.1086/177381},
  \href {http://adsabs.harvard.edu/abs/1996ApJ...464..943I} {464, 943}

\bibitem[\protect\citeauthoryear{{Itoh}, {Adachi}, {Nakagawa}, {Kohyama}  \&
  {Munakata}}{{Itoh} et~al.}{1989}]{Itoh89}
{Itoh} N.,  {Adachi} T.,  {Nakagawa} M.,  {Kohyama} Y.,   {Munakata} H.,  1989,
  \mn@doi [\apj] {10.1086/167301}, \href
  {http://adsabs.harvard.edu/abs/1989ApJ...339..354I} {339, 354}

\bibitem[\protect\citeauthoryear{{Itoh}, {Hayashi}, {Nishikawa}  \&
  {Kohyama}}{{Itoh} et~al.}{1996}]{Itoh96}
{Itoh} N.,  {Hayashi} H.,  {Nishikawa} A.,   {Kohyama} Y.,  1996, \mn@doi
  [\apjs] {10.1086/192264}, \href
  {http://adsabs.harvard.edu/abs/1996ApJS..102..411I} {102, 411}

\bibitem[\protect\citeauthoryear{{Jerkstrand} et~al.,}{{Jerkstrand}
  et~al.}{2016a}]{2016arXiv160802994J}
{Jerkstrand} A.,  et~al., 2016a, preprint, \href
  {http://adsabs.harvard.edu/abs/2016arXiv160802994J} {} (\mn@eprint {arXiv}
  {1608.02994})

\bibitem[\protect\citeauthoryear{{Jerkstrand}, {Smartt}  \&
  {Heger}}{{Jerkstrand} et~al.}{2016b}]{2016MNRAS.455.3207J}
{Jerkstrand} A.,  {Smartt} S.~J.,   {Heger} A.,  2016b, \mn@doi [\mnras]
  {10.1093/mnras/stv2369}, \href
  {http://adsabs.harvard.edu/abs/2016MNRAS.455.3207J} {455, 3207}

\bibitem[\protect\citeauthoryear{{Kasen}, {Woosley}  \& {Heger}}{{Kasen}
  et~al.}{2011}]{2011ApJ...734..102K}
{Kasen} D.,  {Woosley} S.~E.,   {Heger} A.,  2011, \mn@doi [\apj]
  {10.1088/0004-637X/734/2/102}, \href
  {http://adsabs.harvard.edu/abs/2011ApJ...734..102K} {734, 102}

\bibitem[\protect\citeauthoryear{{Kotera}, {Phinney}  \& {Olinto}}{{Kotera}
  et~al.}{2013}]{2013MNRAS.432.3228K}
{Kotera} K.,  {Phinney} E.~S.,   {Olinto} A.~V.,  2013, \mn@doi [\mnras]
  {10.1093/mnras/stt680}, \href
  {http://adsabs.harvard.edu/abs/2013MNRAS.432.3228K} {432, 3228}

\bibitem[\protect\citeauthoryear{{Kozyreva}, {Blinnikov}, {Langer}  \&
  {Yoon}}{{Kozyreva} et~al.}{2014}]{2014A&A...565A..70K}
{Kozyreva} A.,  {Blinnikov} S.,  {Langer} N.,   {Yoon} S.-C.,  2014, \mn@doi
  [\aap] {10.1051/0004-6361/201423447}, \href
  {http://adsabs.harvard.edu/abs/2014A%26A...565A..70K} {565, A70}

\bibitem[\protect\citeauthoryear{{Kozyreva}, {Hirschi}, {Blinnikov}  \& {den
  Hartogh}}{{Kozyreva} et~al.}{2016}]{2016MNRAS.tmpL..20K}
{Kozyreva} A.,  {Hirschi} R.,  {Blinnikov} S.,   {den Hartogh} J.,  2016,
  \mn@doi [\mnras] {10.1093/mnrasl/slw036}, \href
  {http://adsabs.harvard.edu/abs/2016MNRAS.tmpL..20K} {}

\bibitem[\protect\citeauthoryear{{Kurucz} \& {Bell}}{{Kurucz} \&
  {Bell}}{1995}]{1995all..book.....K}
{Kurucz} R.~L.,  {Bell} B.,  1995, {Atomic line list}

\bibitem[\protect\citeauthoryear{{Lada} \& {Lada}}{{Lada} \&
  {Lada}}{2003}]{2003ARA&A..41...57L}
{Lada} C.~J.,  {Lada} E.~A.,  2003, \mn@doi [\araa]
  {10.1146/annurev.astro.41.011802.094844}, \href
  {http://adsabs.harvard.edu/abs/2003ARA%26A..41...57L} {41, 57}

\bibitem[\protect\citeauthoryear{{Langer}, {Norman}, {de Koter}, {Vink},
  {Cantiello}  \& {Yoon}}{{Langer} et~al.}{2007}]{2007A&A...475L..19L}
{Langer} N.,  {Norman} C.~A.,  {de Koter} A.,  {Vink} J.~S.,  {Cantiello} M.,
  {Yoon} S.-C.,  2007, \mn@doi [\aap] {10.1051/0004-6361:20078482}, \href
  {http://adsabs.harvard.edu/abs/2007A%26A...475L..19L} {475, L19}

\bibitem[\protect\citeauthoryear{{Leaman}, {Li}, {Chornock}  \&
  {Filippenko}}{{Leaman} et~al.}{2011}]{2011MNRAS.412.1419L}
{Leaman} J.,  {Li} W.,  {Chornock} R.,   {Filippenko} A.~V.,  2011, \mn@doi
  [\mnras] {10.1111/j.1365-2966.2011.18158.x}, \href
  {http://adsabs.harvard.edu/abs/2011MNRAS.412.1419L} {412, 1419}

\bibitem[\protect\citeauthoryear{{Lee}, {Deane}  \& {Federrath}}{{Lee}
  et~al.}{2009}]{2009ASPC..406..243L}
{Lee} D.,  {Deane} A.~E.,   {Federrath} C.,  2009, in {Pogorelov} N.~V.,
  {Audit} E.,  {Colella} P.,   {Zank} G.~P.,  eds,  Astronomical Society of the
  Pacific Conference Series Vol. 406, Numerical Modeling of Space Plasma Flows:
  ASTRONUM-2008. p.~243

\bibitem[\protect\citeauthoryear{{Li}, {Chornock}, {Leaman}, {Filippenko},
  {Poznanski}, {Wang}, {Ganeshalingam}  \& {Mannucci}}{{Li}
  et~al.}{2011}]{2011MNRAS.412.1473L}
{Li} W.,  {Chornock} R.,  {Leaman} J.,  {Filippenko} A.~V.,  {Poznanski} D.,
  {Wang} X.,  {Ganeshalingam} M.,   {Mannucci} F.,  2011, \mn@doi [\mnras]
  {10.1111/j.1365-2966.2011.18162.x}, \href
  {http://adsabs.harvard.edu/abs/2011MNRAS.412.1473L} {412, 1473}

\bibitem[\protect\citeauthoryear{{Livne}}{{Livne}}{1993}]{1993ApJ...412..634L}
{Livne} E.,  1993, \mn@doi [\apj] {10.1086/172950}, \href
  {http://adsabs.harvard.edu/abs/1993ApJ...412..634L} {412, 634}

\bibitem[\protect\citeauthoryear{{Metzger}, {Vurm}, {Hasco{\"e}t}  \&
  {Beloborodov}}{{Metzger} et~al.}{2014}]{2014MNRAS.437..703M}
{Metzger} B.~D.,  {Vurm} I.,  {Hasco{\"e}t} R.,   {Beloborodov} A.~M.,  2014,
  \mn@doi [\mnras] {10.1093/mnras/stt1922}, \href
  {http://adsabs.harvard.edu/abs/2014MNRAS.437..703M} {437, 703}

\bibitem[\protect\citeauthoryear{{Moriya}}{{Moriya}}{2013}]{2013PhDTMoriya}
{Moriya} T.~J.,  2013, PhD thesis, Department of Astronomy, Graduate School of
  Science University of Tokyo, 206 pp.

\bibitem[\protect\citeauthoryear{{Moriya}, {Tominaga}, {Blinnikov}, {Baklanov}
  \& {Sorokina}}{{Moriya} et~al.}{2011}]{2011MNRAS.415..199M}
{Moriya} T.,  {Tominaga} N.,  {Blinnikov} S.~I.,  {Baklanov} P.~V.,
  {Sorokina} E.~I.,  2011, \mn@doi [\mnras] {10.1111/j.1365-2966.2011.18689.x},
  \href {http://adsabs.harvard.edu/abs/2011MNRAS.415..199M} {415, 199}

\bibitem[\protect\citeauthoryear{{Moriya}, {Pruzhinskaya}, {Ergon}  \&
  {Blinnikov}}{{Moriya} et~al.}{2016}]{2016MNRAS.455..423M}
{Moriya} T.~J.,  {Pruzhinskaya} M.~V.,  {Ergon} M.,   {Blinnikov} S.~I.,  2016,
  \mn@doi [\mnras] {10.1093/mnras/stv2336}, \href
  {http://adsabs.harvard.edu/abs/2016MNRAS.455..423M} {455, 423}

\bibitem[\protect\citeauthoryear{{Nicholl} et~al.,}{{Nicholl}
  et~al.}{2013}]{2013Natur.502..346N}
{Nicholl} M.,  et~al., 2013, \mn@doi [\nat] {10.1038/nature12569}, \href
  {http://adsabs.harvard.edu/abs/2013Natur.502..346N} {502, 346}

\bibitem[\protect\citeauthoryear{{Nicholl} et~al.,}{{Nicholl}
  et~al.}{2014}]{2014MNRAS.444.2096N}
{Nicholl} M.,  et~al., 2014, \mn@doi [\mnras] {10.1093/mnras/stu1579}, \href
  {http://adsabs.harvard.edu/abs/2014MNRAS.444.2096N} {444, 2096}

\bibitem[\protect\citeauthoryear{{Nicholl} et~al.,}{{Nicholl}
  et~al.}{2015}]{2015MNRAS.452.3869N}
{Nicholl} M.,  et~al., 2015, \mn@doi [\mnras] {10.1093/mnras/stv1522}, \href
  {http://adsabs.harvard.edu/abs/2015MNRAS.452.3869N} {452, 3869}

\bibitem[\protect\citeauthoryear{{Noebauer} \& {Sim}}{{Noebauer} \&
  {Sim}}{2015}]{2015MNRAS.453.3120N}
{Noebauer} U.~M.,  {Sim} S.~A.,  2015, \mn@doi [\mnras]
  {10.1093/mnras/stv1849}, \href
  {http://adsabs.harvard.edu/abs/2015MNRAS.453.3120N} {453, 3120}

\bibitem[\protect\citeauthoryear{{Noebauer}, {Sim}, {Kromer}, {R{\"o}pke}  \&
  {Hillebrandt}}{{Noebauer} et~al.}{2012}]{2012MNRAS.425.1430N}
{Noebauer} U.~M.,  {Sim} S.~A.,  {Kromer} M.,  {R{\"o}pke} F.~K.,
  {Hillebrandt} W.,  2012, \mn@doi [\mnras] {10.1111/j.1365-2966.2012.21600.x},
  \href {http://adsabs.harvard.edu/abs/2012MNRAS.425.1430N} {425, 1430}

\bibitem[\protect\citeauthoryear{{Nomoto}, {Thielemann}  \& {Yokoi}}{{Nomoto}
  et~al.}{1984}]{1984ApJ...286..644N}
{Nomoto} K.,  {Thielemann} F.-K.,   {Yokoi} K.,  1984, \mn@doi [\apj]
  {10.1086/162639}, \href {http://adsabs.harvard.edu/abs/1984ApJ...286..644N}
  {286, 644}

\bibitem[\protect\citeauthoryear{{Nugis} \& {Lamers}}{{Nugis} \&
  {Lamers}}{2000}]{2000A&A...360..227N}
{Nugis} T.,  {Lamers} H.~J.~G.~L.~M.,  2000, \aap, \href
  {http://adsabs.harvard.edu/abs/2000A%26A...360..227N} {360, 227}

\bibitem[\protect\citeauthoryear{{Pan}, {Kasen}  \& {Loeb}}{{Pan}
  et~al.}{2012}]{2012MNRAS.422.2701P}
{Pan} T.,  {Kasen} D.,   {Loeb} A.,  2012, \mn@doi [\mnras]
  {10.1111/j.1365-2966.2012.20837.x}, \href
  {http://adsabs.harvard.edu/abs/2012MNRAS.422.2701P} {422, 2701}

\bibitem[\protect\citeauthoryear{{Phillips} et~al.,}{{Phillips}
  et~al.}{2007}]{2007PASP..119..360P}
{Phillips} M.~M.,  et~al., 2007, \mn@doi [\pasp] {10.1086/518372}, \href
  {http://adsabs.harvard.edu/abs/2007PASP..119..360P} {119, 360}

\bibitem[\protect\citeauthoryear{{Quimby} et~al.,}{{Quimby}
  et~al.}{2011}]{2011Natur.474..487Q}
{Quimby} R.~M.,  et~al., 2011, \mn@doi [\nat] {10.1038/nature10095}, \href
  {http://adsabs.harvard.edu/abs/2011Natur.474..487Q} {474, 487}

\bibitem[\protect\citeauthoryear{{Rakavy} \& {Shaviv}}{{Rakavy} \&
  {Shaviv}}{1967}]{1967ApJ...148..803R}
{Rakavy} G.,  {Shaviv} G.,  1967, \mn@doi [\apj] {10.1086/149204}, \href
  {http://adsabs.harvard.edu/abs/1967ApJ...148..803R} {148, 803}

\bibitem[\protect\citeauthoryear{{Richardson}, {Jenkins}, {Wright}  \&
  {Maddox}}{{Richardson} et~al.}{2014}]{2014AJ....147..118R}
{Richardson} D.,  {Jenkins} III R.~L.,  {Wright} J.,   {Maddox} L.,  2014,
  \mn@doi [\aj] {10.1088/0004-6256/147/5/118}, \href
  {http://adsabs.harvard.edu/abs/2014AJ....147..118R} {147, 118}

\bibitem[\protect\citeauthoryear{{Roth} \& {Kasen}}{{Roth} \&
  {Kasen}}{2015}]{2015ApJS..217....9R}
{Roth} N.,  {Kasen} D.,  2015, \mn@doi [\apjs] {10.1088/0067-0049/217/1/9},
  \href {http://adsabs.harvard.edu/abs/2015ApJS..217....9R} {217, 9}

\bibitem[\protect\citeauthoryear{{Rydberg}, {Zackrisson}, {Lundqvist}  \&
  {Scott}}{{Rydberg} et~al.}{2013}]{2013MNRAS.429.3658R}
{Rydberg} C.-E.,  {Zackrisson} E.,  {Lundqvist} P.,   {Scott} P.,  2013,
  \mn@doi [\mnras] {10.1093/mnras/sts653}, \href
  {http://adsabs.harvard.edu/abs/2013MNRAS.429.3658R} {429, 3658}

\bibitem[\protect\citeauthoryear{{Scannapieco}, {Madau}, {Woosley}, {Heger}  \&
  {Ferrara}}{{Scannapieco} et~al.}{2005}]{2005ApJ...633.1031S}
{Scannapieco} E.,  {Madau} P.,  {Woosley} S.,  {Heger} A.,   {Ferrara} A.,
  2005, \mn@doi [\apj] {10.1086/444450}, \href
  {http://adsabs.harvard.edu/abs/2005ApJ...633.1031S} {633, 1031}

\bibitem[\protect\citeauthoryear{{Schneider} et~al.,}{{Schneider}
  et~al.}{2014}]{2014ApJ...780..117S}
{Schneider} F.~R.~N.,  et~al., 2014, \mn@doi [\apj]
  {10.1088/0004-637X/780/2/117}, \href
  {http://adsabs.harvard.edu/abs/2014ApJ...780..117S} {780, 117}

\bibitem[\protect\citeauthoryear{{Sorokina}, {Blinnikov}  \&
  {Bartunov}}{{Sorokina} et~al.}{2000}]{2000AstL...26...67S}
{Sorokina} E.~I.,  {Blinnikov} S.~I.,   {Bartunov} O.~S.,  2000, \mn@doi
  [Astronomy Letters] {10.1134/1.20370}, \href
  {http://adsabs.harvard.edu/abs/2000AstL...26...67S} {26, 67}

\bibitem[\protect\citeauthoryear{{Sorokina}, {Blinnikov}, {Nomoto}, {Quimby}
  \& {Tolstov}}{{Sorokina} et~al.}{2015}]{2015arXiv151000834S}
{Sorokina} E.,  {Blinnikov} S.,  {Nomoto} K.,  {Quimby} R.,   {Tolstov} A.,
  2015, preprint, \href {http://adsabs.harvard.edu/abs/2015arXiv151000834S} {}
  (\mn@eprint {arXiv} {1510.00834})

\bibitem[\protect\citeauthoryear{{Stabrowski}}{{Stabrowski}}{1997}]{Stabrowski%
1997}
{Stabrowski} M.~M.,  1997, \mn@doi [Simulation Practice and Theory]
  {http://dx.doi.org/10.1016/S0928-4869(96)00011-0}, 5, 333

\bibitem[\protect\citeauthoryear{{Swartz}, {Sutherland}  \&
  {Harkness}}{{Swartz} et~al.}{1995}]{1995ApJ...446..766S}
{Swartz} D.~A.,  {Sutherland} P.~G.,   {Harkness} R.~P.,  1995, \mn@doi [\apj]
  {10.1086/175834}, \href {http://adsabs.harvard.edu/abs/1995ApJ...446..766S}
  {446, 766}

\bibitem[\protect\citeauthoryear{{Th{\"o}ne}, {de Ugarte Postigo},
  {Garc{\'{\i}}a-Benito}, {Leloudas}, {Schulze}  \&
  {Amor{\'{\i}}n}}{{Th{\"o}ne} et~al.}{2015}]{2015MNRAS.451L..65T}
{Th{\"o}ne} C.~C.,  {de Ugarte Postigo} A.,  {Garc{\'{\i}}a-Benito} R.,
  {Leloudas} G.,  {Schulze} S.,   {Amor{\'{\i}}n} R.,  2015, \mn@doi [\mnras]
  {10.1093/mnrasl/slv051}, \href
  {http://adsabs.harvard.edu/abs/2015MNRAS.451L..65T} {451, L65}

\bibitem[\protect\citeauthoryear{{Timmes} \& {Swesty}}{{Timmes} \&
  {Swesty}}{2000}]{2000ApJS..126..501T}
{Timmes} F.~X.,  {Swesty} F.~D.,  2000, \mn@doi [\apjs] {10.1086/313304}, \href
  {http://adsabs.harvard.edu/abs/2000ApJS..126..501T} {126, 501}

\bibitem[\protect\citeauthoryear{{Tolstov}, {Nomoto}  \& {Blinnikov}}{{Tolstov}
  et~al.}{2016a}]{2016Tolstov}
{Tolstov} A.~G.,  {Nomoto} K.,   {Blinnikov} S.~I.,  2016a, submitted to ApJ

\bibitem[\protect\citeauthoryear{{Tolstov}, {Nomoto}, {Tominaga}, {Ishigaki},
  {Blinnikov}  \& {Suzuki}}{{Tolstov} et~al.}{2016b}]{2016ApJ...821..124T}
{Tolstov} A.,  {Nomoto} K.,  {Tominaga} N.,  {Ishigaki} M.~N.,  {Blinnikov} S.,
    {Suzuki} T.,  2016b, \mn@doi [\apj] {10.3847/0004-637X/821/2/124}, \href
  {http://adsabs.harvard.edu/abs/2016ApJ...821..124T} {821, 124}

\bibitem[\protect\citeauthoryear{{Verner}, {Verner}  \& {Ferland}}{{Verner}
  et~al.}{1996}]{1996ADNDT..64....1V}
{Verner} D.~A.,  {Verner} E.~M.,   {Ferland} G.~J.,  1996, \mn@doi [Atomic Data
  and Nuclear Data Tables] {10.1006/adnd.1996.0018}, \href
  {http://adsabs.harvard.edu/abs/1996ADNDT..64....1V} {64, 1}

\bibitem[\protect\citeauthoryear{{Vink}}{{Vink}}{2015}]{2015ASSL..412...77V}
{Vink} J.~S.,  2015, in {Vink} J.~S.,  ed.,  Astrophysics and Space Science
  Library Vol. 412, Very Massive Stars in the Local Universe. p.~77 (\mn@eprint
  {arXiv} {1406.5357}), \mn@doi{10.1007/978-3-319-09596-7_4}

\bibitem[\protect\citeauthoryear{{Vink}, {de Koter}  \& {Lamers}}{{Vink}
  et~al.}{2001}]{2001A&A...369..574V}
{Vink} J.~S.,  {de Koter} A.,   {Lamers} H.~J.~G.~L.~M.,  2001, \mn@doi [\aap]
  {10.1051/0004-6361:20010127}, \href
  {http://adsabs.harvard.edu/abs/2001A%26A...369..574V} {369, 574}

\bibitem[\protect\citeauthoryear{{Vreeswijk} et~al.,}{{Vreeswijk}
  et~al.}{2016}]{2016arXiv160908145V}
{Vreeswijk} P.~M.,  et~al., 2016, preprint, \href
  {http://adsabs.harvard.edu/abs/2016arXiv160908145V} {} (\mn@eprint {arXiv}
  {1609.08145})

\bibitem[\protect\citeauthoryear{{Weaver}, {Zimmerman}  \& {Woosley}}{{Weaver}
  et~al.}{1978}]{1978ApJ...225.1021W}
{Weaver} T.~A.,  {Zimmerman} G.~B.,   {Woosley} S.~E.,  1978, \mn@doi [\apj]
  {10.1086/156569}, \href {http://adsabs.harvard.edu/abs/1978ApJ...225.1021W}
  {225, 1021}

\bibitem[\protect\citeauthoryear{{Whalen}, {Fryer}, {Holz}, {Heger}, {Woosley},
  {Stiavelli}, {Even}  \& {Frey}}{{Whalen} et~al.}{2013a}]{2013ApJ...762L...6W}
{Whalen} D.~J.,  {Fryer} C.~L.,  {Holz} D.~E.,  {Heger} A.,  {Woosley} S.~E.,
  {Stiavelli} M.,  {Even} W.,   {Frey} L.~H.,  2013a, \mn@doi [\apjl]
  {10.1088/2041-8205/762/1/L6}, \href
  {http://adsabs.harvard.edu/abs/2013ApJ...762L...6W} {762, L6}

\bibitem[\protect\citeauthoryear{{Whalen} et~al.,}{{Whalen}
  et~al.}{2013b}]{2013ApJ...777..110W}
{Whalen} D.~J.,  et~al., 2013b, \mn@doi [\apj] {10.1088/0004-637X/777/2/110},
  \href {http://adsabs.harvard.edu/abs/2013ApJ...777..110W} {777, 110}

\bibitem[\protect\citeauthoryear{{Wollaeger} \& {van Rossum}}{{Wollaeger} \&
  {van Rossum}}{2014}]{2014ApJS..214...28W}
{Wollaeger} R.~T.,  {van Rossum} D.~R.,  2014, \mn@doi [\apjs]
  {10.1088/0067-0049/214/2/28}, \href
  {http://adsabs.harvard.edu/abs/2014ApJS..214...28W} {214, 28}

\bibitem[\protect\citeauthoryear{{Wollaeger}, {van Rossum}, {Graziani},
  {Couch}, {Jordan}, {Lamb}  \& {Moses}}{{Wollaeger}
  et~al.}{2013}]{2013ApJS..209...36W}
{Wollaeger} R.~T.,  {van Rossum} D.~R.,  {Graziani} C.,  {Couch} S.~M.,
  {Jordan} IV G.~C.,  {Lamb} D.~Q.,   {Moses} G.~A.,  2013, \mn@doi [\apjs]
  {10.1088/0067-0049/209/2/36}, \href
  {http://adsabs.harvard.edu/abs/2013ApJS..209...36W} {209, 36}

\bibitem[\protect\citeauthoryear{{Woodward} \& {Colella}}{{Woodward} \&
  {Colella}}{1984}]{1984JCoPh..54..115W}
{Woodward} P.,  {Colella} P.,  1984, \mn@doi [Journal of Computational Physics]
  {10.1016/0021-9991(84)90142-6}, \href
  {http://adsabs.harvard.edu/abs/1984JCoPh..54..115W} {54, 115}

\bibitem[\protect\citeauthoryear{{Woosley} \& {Heger}}{{Woosley} \&
  {Heger}}{2015}]{2015ASSL..412..199W}
{Woosley} S.~E.,  {Heger} A.,  2015, in {Vink} J.~S.,  ed.,  Astrophysics and
  Space Science Library Vol. 412, Very Massive Stars in the Local Universe.
  p.~199 (\mn@eprint {arXiv} {1406.5657}), \mn@doi{10.1007/978-3-319-09596-7_7}

\bibitem[\protect\citeauthoryear{{Woosley}, {Heger}  \& {Weaver}}{{Woosley}
  et~al.}{2002}]{2002RvMP...74.1015W}
{Woosley} S.~E.,  {Heger} A.,   {Weaver} T.~A.,  2002, \mn@doi [Reviews of
  Modern Physics] {10.1103/RevModPhys.74.1015}, \href
  {http://adsabs.harvard.edu/abs/2002RvMP...74.1015W} {74, 1015}

\bibitem[\protect\citeauthoryear{{Woosley}, {Kasen}, {Blinnikov}  \&
  {Sorokina}}{{Woosley} et~al.}{2007}]{2007ApJ...662..487W}
{Woosley} S.~E.,  {Kasen} D.,  {Blinnikov} S.,   {Sorokina} E.,  2007, \mn@doi
  [\apj] {10.1086/513732}, \href
  {http://adsabs.harvard.edu/abs/2007ApJ...662..487W} {662, 487}

\bibitem[\protect\citeauthoryear{{Yusof} et~al.,}{{Yusof}
  et~al.}{2013}]{2013MNRAS.433.1114Y}
{Yusof} N.,  et~al., 2013, \mn@doi [\mnras] {10.1093/mnras/stt794}, \href
  {http://adsabs.harvard.edu/abs/2013MNRAS.433.1114Y} {433, 1114}

\bibitem[\protect\citeauthoryear{{Zeldovich} \& {Novikov}}{{Zeldovich} \&
  {Novikov}}{1971}]{1971reas.book.....Z}
{Zeldovich} Y.~B.,  {Novikov} I.~D.,  1971, {Relativistic astrophysics. Vol.1:
  Stars and relativity}.
Chicago: University of Chicago Press, Translated by Eli Arlock, Ed. by
  K.S.~Thorne and W.D.~Arnett

\bibitem[\protect\citeauthoryear{{Zel'dovich}, {Blinnikov}  \&
  {Shakura}}{{Zel'dovich} et~al.}{1981}]{1981ppse.book.....Z}
{Zel'dovich} Y.~B.,  {Blinnikov} S.~I.,   {Shakura} N.~I.,  1981, {Physical
  principles of structure and evolution of stars}

\bibitem[\protect\citeauthoryear{{Zinnecker} \& {Yorke}}{{Zinnecker} \&
  {Yorke}}{2007}]{2007ARA&A..45..481Z}
{Zinnecker} H.,  {Yorke} H.~W.,  2007, \mn@doi [\araa]
  {10.1146/annurev.astro.44.051905.092549}, \href
  {http://adsabs.harvard.edu/abs/2007ARA%26A..45..481Z} {45, 481}

\bibitem[\protect\citeauthoryear{{de Jager}, {Nieuwenhuijzen}  \& {van der
  Hucht}}{{de Jager} et~al.}{1988}]{deJager88}
{de Jager} C.,  {Nieuwenhuijzen} H.,   {van der Hucht} K.~A.,  1988, \aaps,
  \href {http://adsabs.harvard.edu/abs/1988A%26AS...72..259D} {72, 259}

\bibitem[\protect\citeauthoryear{{von Neumann} \& {Richtmyer}}{{von Neumann} \&
  {Richtmyer}}{1950}]{1950vonNeumann}
{von Neumann} J.,  {Richtmyer} R.~D.,  1950, \mn@doi [J. Appl. Phys. 21, 232]
  {10.1063/1.1699639}, 21, 232

\makeatother
\end{thebibliography}

\appendix
\section[Description of radiation codes used for comparison analysis]
{Description of radiation codes used for comparison analysis}
\label{sect:append1}

%
%

\subsection[MCRH]{MCRH}
\label{subsect:MCRH}

\texttt{MCRH} \citep{2012MNRAS.425.1430N,2015MNRAS.453.3120N} is a
one-dimensional Monte-Carlo (MC) radiation hydrodynamics code.  Light curves for
supernova ejecta are computed adopting the following assumptions:
\begin{itemize}
\item radiative equilibrium: all radiation--matter interactions are treated as pure scatterings;
\item radiation--matter interactions only transfer momentum
but do not affect the internal energy balance;
\item $\gamma$-rays generated in the nickel-56 and cobalt-56 decay are tracked in a separate MC step;
 their interactions with the medium are described by a grey pure absorption cross section $\kappa = 0.03 \,
 \mathrm{cm^{\,2}\,s^{\,-1}}$;
  Once a $\gamma$-ray photon is absorbed, it is instantaneously converted into radiation energy (which is tracked by the main MC routine).
\item in contrast to the SNe~Ia calculations presented by
  \citet{2012MNRAS.425.1430N}, a constant radiation scattering cross-section is
  used (either $0.05$ or $0.1\,\mathrm{cm^{\,2}\,g^{\,-1}}$).
\end{itemize}
 
The \texttt{MCRH} simulations are started at day~10 after the explosion.  
The phase prior to the starting point is treated in an analytic homologous
expansion of the \texttt{STELLA} profile at day~1 according to the
following relations:
\begin{equation}
r=r_0 \left({t\over t_0}\right) \quad \rho = \rho_{\,0} \left({t\over t_0}\right)^{\,-3}\, .
\label{equation:homology}
\end{equation}
Between day~1 and day~10, the decay of nickel-56 is taken into account and the
released energy is tracked.  This energy, together with the initial thermal
field, is used to set the radiation field at the beginning of the calculation
and after accounting for adiabatic cooling losses.  For the \texttt{MCRH}
simulations, the outermost cells of the input \texttt{STELLA} profile are
discarded (with $v > 0.1\,c$), since the high velocities in these regions are
incompatible with the current design of \texttt{MCRH}, which only takes
relativistic terms of $\mathcal{O}(v/c)$ into account.  
In all \texttt{MCRH} calculations presented here, the initial radiation field is
discretized by 100,000 MC packets.

\subsection[SuperNu]{SuperNu}
\label{subsect:SuperNu}

\texttt{SuperNu} is a multigroup LTE radiative transfer code that employs Implicit
Monte Carlo (IMC) and Discrete Diffusion Monte Carlo 
\citep[DDMC,][van Rossum, in preparation]{2013ApJS..209...36W,2014ApJS..214...28W}.  
IMC solves the thermal radiative transfer equations semi-implicitly by
treating some absorption and emission as instantaneous effective scattering
\citep[see, e.g.,][]{fleck1971}.  Thus even in purely absorbing media,
MC particles can undergo isotropic scattering and wavelength redistribution.  
DDMC accelerates IMC over optically thick regions of space
\citep{densmore2007} and ranges of wavelength
\citep{densmore2012,abdikamalov2012} by replacing many low mean-free-path
scattering events with single leakage events.  
\texttt{SuperNu} can apply IMC and DDMC in both static and homologous,
semi-relativistically expanding atmospheres.  
The code has been verified by analytic and semi-analytic radiative transfer
tests \citep{2013ApJS..209...36W} and on the W7 model of SNe~Ia
\citep{1984ApJ...286..644N,2014ApJS..214...28W}.  


For the constant-opacity test, \texttt{SuperNu} was started at day~10
with the same setup as \texttt{MCRH} simulations (as described in
Section~\ref{subsect:MCRH}).  For the gamma-ray
transfer, \texttt{SuperNu} employed a constant absorption opacity of
0.03~cm$^2$\,g$^{\,-1}$ as in \texttt{MCRH}.  
The gamma-ray packets in \texttt{SuperNu} are not directly converted to optical
packets, but instead are used to tally the total gamma-ray energy deposition
per spatial cell.  The deposition energy values are then added to the
thermal source for optical packets.

\subsection[V1D]{V1D}
\label{subsect:V1D}

\texttt{V1D} is a one-dimentional hydrodynamics version of the code
\texttt{Vulkan} \citep{1993ApJ...412..634L}.  
\texttt{V1D} solves the equations of motion using explicit Lagrangian
hydrodynamics, implicitly coupled with the equations of radiative transfer.  The
radiative-transport is solved under the approximations of LTE and grey
diffusion.  The grey opacities in \texttt{V1D} were computed based on the opacity
routines of \texttt{CMFGEN} 
\citep{2010MNRAS.405.2141D,2010MNRAS.405.2113D,2015MNRAS.449.4304D}.  
Hence, \texttt{V1D} calculates supernova ejecta evolution with coupled
hydrodynamics and radiation.

\bsp	
\label{lastpage}
\end{document}